\documentclass{aa}
\usepackage{graphicx}
\usepackage{txfonts}
\usepackage{natbib}
\usepackage{hyperref}
\usepackage{xcolor}
\usepackage{orcidlink}
\usepackage{subcaption}   
\usepackage{lipsum}
\usepackage[utf8]{inputenc}

\usepackage{siunitx}

\begin{document}

\title{Multilayer magnetic structure toward NGC~1333 revealed by a square-degree optical polarimetric survey}

\author{G.~A.~P.~Franco\inst{1} \orcidlink{0000-0003-2020-2649}
        \and J.~M.~Girart\inst{2,3} \orcidlink{0000-0002-3829-5591}
        \and F.~O.~Alves\inst{4} \orcidlink{0000-0002-7945-064X}}

\institute{Departamento de F\'\i sica -- ICEx -- UFMG, Av. Ant\^onio Carlos 6627, 30.270-901 Belo Horizonte, Brazil 
\and Institut de Ci\`encies de l'Espai (ICE--CSIC), Campus UAB, Carrer de Can Magrans s/n, 08193 Cerdanyola del Vall\`es, 
Catalonia, Spain 
\and Institut d'Estudis Espacials de Catalunya (IEEC), Esteve Terradas 1, Campus del Baix Llobregat – UPC, 08860 Castelldefels, Catalonia, Spain
\and Institut de Radioastronomie Millim\'etrique (IRAM), 300 rue de la Piscine, 38406 Saint-Martin d'H\`eres, France}

\date{Received ; accepted }

\abstract
{The interpretation of magnetic-field morphologies in star-forming regions can be complicated by the superposition 
of interstellar structures at different distances along the line of sight. Stellar distances combined with optical 
polarimetry provide a means of identifying and estimating the contribution of foreground polarization.}
{We investigated the large-scale magnetic-field morphology toward NGC~1333 in the Perseus molecular cloud, 
aiming to characterize the field associated with the cloud while accounting for the contribution of an intervening 
polarized layer.}
{We conducted a wide-field $R$-band optical polarimetric survey covering nearly one square degree toward 
NGC~1333, obtaining linear polarization measurements for approximately 840 stars with distance estimates 
from \textit{Gaia} DR3 and \texttt{StarHorse}. We analyzed the polarization--distance relation and the Stokes parameters 
to identify distinct line-of-sight components and estimate the contribution of an intermediate-distance foreground layer. 
After subtracting this contribution in Stokes space, we characterized the magnetic-field orientation using circular statistics 
and the angular dispersion function (ADF). We also compared the resulting morphology with CO molecular-line emission and 
polarization measurements from \textit{Planck}, near-infrared observations, and JCMT/POL-2.}
{The polarization--distance relation distinguishes a weakly polarizing nearby component at $d\lesssim160$~pc,
an intermediate-distance polarized contribution traced by stars at approximately 160 and 300~pc,
and a component associated with the Perseus molecular cloud at distances 
beyond approximately 300~pc. After foreground correction, the plane-of-sky magnetic field exhibits a smooth large-scale 
rotation across the surveyed region, from a mean orientation of $126\fdg3\pm0\fdg8$ in the northwestern region to $
162\fdg2\pm0\fdg6$ in the southeastern region. Circular statistics indicate that the field remains well ordered in both 
regions. The ADF analysis yields turbulent-to-ordered magnetic-field strength ratios of $0.054^{+0.025}_{-0.022}$ and 
$0.095^{+0.021}_{-0.047}$ in the northwestern and southeastern regions, respectively. The corresponding plane-of-sky 
magnetic-field strengths, $59^{+44}_{-23}~\mu\mathrm{G}$ and $46^{+50}_{-13}~\mu\mathrm{G}$, are consistent within 
their uncertainties. The turbulent-to-ordered magnetic-field ratios and the corresponding plane-of-sky magnetic-field 
strengths should be regarded as approximate estimates, since residual sightline-to-sightline foreground fluctuations 
may contribute to the measured angular dispersion. CO observations reveal the same velocity components in both 
regions but with systematically varying  relative contributions across the field. The foreground-corrected optical 
polarization agrees with the large-scale orientation inferred from \textit{Planck} and is broadly consistent with 
near-infrared and JCMT/POL-2 measurements. Nonetheless, the  latter reveal localized departures in the denser 
star-forming structures.}
{The combination of stellar distances and wide-field optical polarimetry provides a way to estimate foreground polarization 
and investigate the magnetic-field morphology associated with NGC~1333. The results favor a coherent cloud-scale magnetic 
field whose projected orientation changes smoothly across the region, accompanied by systematic variations in the relative 
contributions of molecular-gas velocity components. Comparisons with complementary polarimetric tracers further show that 
this large-scale organization coexists with localized magnetic-field variations within the denser structures.}


\maketitle
 \nolinenumbers

\section{Introduction}

Magnetic fields are an important component of the interstellar medium
and can influence the evolution of molecular clouds, the fragmentation
of dense structures, the transport of angular momentum, and the
formation of protostellar outflows and jets
\citep{Maury:2022, Pattle:2023}. Their role, however, depends on the
physical scale considered. In nearby star-forming regions, magnetic
fields are observed across a wide range of environments, from diffuse
material surrounding molecular clouds to dense filaments and
protostellar cores. Establishing how these different components are
related is particularly challenging when several structures overlap
along the line of sight.

NGC~1333 is one of the most active sites of nearby low-mass star
formation and is located in the western part of the Perseus molecular
cloud complex, a large system of molecular clouds extending over
several degrees on the sky \citep{Walawender:2008}. The region contains
a rich population of young stellar objects, numerous Herbig--Haro
objects, and powerful molecular outflows
\citep[e.g.,][]{Herbig:1983, Arce:2010, Winston:2009}. Its young stellar
population and proximity make NGC~1333 an excellent laboratory for
investigating the coupled evolution of gas, magnetic fields, and newly
formed stars. At the same time, NGC~1333 is part of the dynamically and
spatially structured Perseus complex rather than an isolated molecular
cloud. Gaia-based studies have revealed multiple stellar groups across Perseus with
distinct spatial and kinematic properties. These groups span ages
of several megayears and show evidence of a hierarchical organization of
star formation \citep{Olivares:2023}. More recent analyses of young
stellar objects have further reported large-scale coherence between
stellar motions and the molecular-gas velocity field, demonstrating
that the stellar and gaseous components of Perseus retain signatures of
their common dynamical environment \citep{Gu:2026}.

The 3D structure of Perseus is also important for the
interpretation of magnetic-field observations. Distance estimates to
the complex have evolved considerably, from early suggestions of
multiple dust layers along the line of sight
\citep{Cernis:1990, Cernis:1993} to trigonometric distances based on
water masers \citep{Hirota:2008} and subsequent photometric and
\textit{Gaia}-based determinations that reveal distance gradients
across the cloud \citep{Schlafly:2014, Ortiz:2018, Zucker:2018}.
In particular, the line of sight toward Perseus contains material at
different distances, so the optical polarization measured toward a
given star need not originate solely in the molecular cloud associated
with NGC~1333. Distance-resolved stellar polarimetry has indeed shown
that the Perseus sight lines contain more than one magnetized component.
For example, \citet{Doi:2021} identify distinct polarization
components associated with foreground and more distant material, with
characteristic distances of order $\sim$150 and $\sim$300 pc,
respectively. These results established the importance of resolving the
line-of-sight structure before interpreting stellar polarization as a
direct tracer of the magnetic field of the molecular cloud itself
\citep{Doi:2021}.

The magnetic field of Perseus and NGC~1333 has been investigated over
several different physical scales. Early optical polarimetric surveys
covered large areas of the Perseus complex and revealed complex
polarization-angle distributions, including evidence of multiple
components in the plane of the sky \citep{Vrba:1976, Goodman:1990}.
These wide-field optical measurements primarily probe the more diffuse
material along extended lines of sight. By contrast, near-infrared
polarimetric observations of the IRAS~4A region focused on much smaller
spatial scales and on material immediately surrounding a protostellar
system \citep{Tamura:1988, Alves:2011}. At the dense-core and protostellar scales, interferometric
polarimetry has resolved magnetic-field structures, which cannot be
captured by cloud-scale surveys. SMA observations of NGC~1333 IRAS~4A
revealed an hourglass-shaped field on scales of a few hundred AU
\citep{Girart:2006}, while subsequent ALMA and JVLA observations
examined the polarization morphology from the submillimeter to
centimeter wavelengths at scales of approximately 100--1000 AU
\citep{Ko:2020}. On the larger scales of dense filaments, submillimeter
polarimetry with \textit{SCUBA-2}/POL-2 mapped the field in NGC~1333
\citep{Doi:2020}, and a later study examined the internal magnetic
structure of individual filaments \citep{Doi:2021a}. Taken together,
these studies show that field morphology depends on both the spatial scale
and the density regime being traced.

The large-scale diffuse field is also probed by space-based
submillimeter observations. \textit{Planck} measurements provide the
magnetic-field orientation over much broader areas and trace dust
emission from relatively diffuse material compared with the dense
cores mapped by ground-based interferometric and submillimeter
observations \citep{Planck_XXXII:2016}. The comparison between these
datasets should therefore be interpreted in terms of their different
angular resolutions, column-density regimes, and line-of-sight
weighting rather than as measurements of a single magnetic component.
Together, the existing observations show that NGC~1333 contains
magnetic-field information spanning several physical scales, but they
do not by themselves provide a continuous view of how the field in the
diffuse material surrounding the dense structures connects to the
cloud-scale magnetic morphology.

The dynamical environment of NGC~1333 may further contribute to this
complexity. The cloud contains multiple velocity components and a
highly fragmented filamentary network
\citep{Hacar:2017, Dhabal:2018, Dhabal:2019}. High-resolution
observations of the southern part of NGC~1333 have revealed
shock-related structures near IRAS~4A and have suggested that external
perturbations may influence the surrounding molecular gas
\citep{DeSimone:2022}. In particular, \citet{DeSimone:2022} report a
change in the orientation of the large-scale shocked structures that
appears to be accompanied by a change in the local magnetic-field
orientation. Although the physical origin of these perturbations and
their relationship to the cloud-scale field remain uncertain, such
results illustrate that magnetic-field morphology in NGC~1333 may
reflect the combined effects of cloud structure, gas dynamics, and
localized perturbations.

These previous studies motivate a wide-field stellar polarimetric
approach that explicitly uses distance information to separate
foreground and cloud contributions. The principal novelty of the
present work is not the identification of a foreground polarization
component itself, which is consistent with earlier distance-resolved
studies, but the characterization of that component in the specific
field surrounding NGC~1333 and the use of the resulting corrected
stellar polarization sample to investigate the magnetic field on
scales intermediate between the diffuse cloud environment and the
dense structures previously mapped at high angular resolution.

We present a wide-field optical polarimetric survey covering nearly
one square degree around NGC~1333, with reliable linear polarization
measurements for more than 800 stars combined with \textit{Gaia}
DR3/\texttt{StarHorse} distance information. These data allow us to
identify the distance range associated with the foreground
polarization and remove its contribution from the stellar Stokes
parameters before analyzing the magnetic field of the Perseus
cloud. We then characterize the spatial organization and variation of
the plane-of-the-sky magnetic field using circular statistics and the
angular dispersion function (ADF) and investigate its connection with the
molecular-gas kinematics and with magnetic-field orientations traced
by infrared and submillimeter observations. By extending the
polarimetric information into the lower-density material surrounding
the dense NGC~1333 structures, our survey provides a complementary
view of the cloud-scale magnetic field and allows its spatial and
angular-scale variations to be examined across several parsecs.

\section{Observations and data reduction}

\subsection{Polarimetric observations}

Observations were conducted with the 1.6\,m telescope at the Observatório do Pico dos 
Dias (LNA/MCT, Brazil) during observing runs in 2007 and 2008. These data were obtained 
with the use of a specially adapted CCD camera to enable polarimetric measurements --- for a 
suitable description of the polarimeter see \citet{Magalhaes:1996}. $R$-band linear polarimetry was 
obtained for 24 fields (with field of view of about $12' \times 12'$ each) almost covering a 
square degree of the area of the sky centered in NGC~1333. 

When in linear polarization mode, the polarimeter incorporates a rotatable, 
achromatic half-wave retarder followed by a calcite Savart plate. The 
half-wave retarder can be rotated in steps of 22\fdg 5, and one polarization 
modulation cycle is covered for every 90\degr\ rotation of this wave plate. 
This arrangement provides two images of each object on the CCD with
perpendicular polarizations (the ordinary, $f_o$, and the extraordinary,
$f_e$, beams). Rotating the half-wave plate by 45\degr\ yields in a rotation
of the polarization direction of 90\degr . Thus, at the CCD area where $f_o$
was first detected, now $f_e$ is imaged and vice versa. Combining all four
intensities reduces flat-field irregularities. In addition, the simultaneous
imaging of the two beams enables observations under non-photometric conditions
and, at the same time, the sky polarization is practically canceled. Eight CCD
images were taken for each field with the polarizer rotated through two
modulation cycles of 0\degr, 22\fdg 5, 45\degr, and 67\fdg 5 in rotation
angle. The integration exposure for each CCD frame, of each position of the 
half-wave plate analyzer, was set to 120 seconds.

\subsection{Data reduction and calibration}

\begin{figure}
\centering
\includegraphics[width=.95\hsize]{{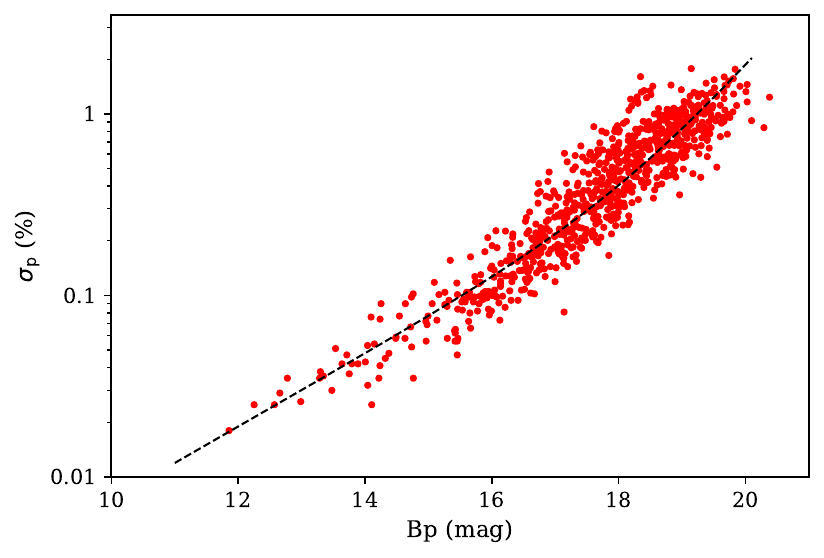}}
\caption{Distribution of the polarimetric errors as a function of the stellar $B_P$ 
magnitude, as retrieved from the \textit{Gaia} DR3 catalog. The red points represent 
the measured uncertainties for individual sources, while the dashed black curve shows 
the best-fit noise model. The fit combines a systematic floor, shot-noise, and sky-noise 
contributions. 
}
\label{fig:sigma_pol}
\end{figure}

Standard CCD reduction procedures (bias subtraction and flat-fielding) were applied, and aperture 
photometry of the ordinary and extraordinary beams was performed using \textsc{IRAF/DAOPHOT}. 
The normalized Stokes parameters $Q$ and $U$ were obtained using the PCCDPACK package 
\citep{Pereyra:2000}, by fitting the normalized intensity differences to the modulation function
\begin{equation}
z_i=Q\cos(4\psi_i)+U\sin(4\psi_i),
\end{equation}
where $\psi_i$ is the half-wave plate position angle for the $i$-th exposure. The degree of linear 
polarization and position angle were calculated as $P=\sqrt{Q^2+U^2}$ and 
$\theta=\frac{1}{2}\operatorname{atan2}(U,Q)$, respectively, with $\theta$ measured from north 
to east. Empirical uncertainties were estimated from the fit residuals and adopted in the 
subsequent analysis. Figure~\ref{fig:sigma_pol} shows the resulting polarimetric uncertainties as a 
function of stellar $G_{\rm BP}$ magnitude from the \textit{Gaia} DR3 catalog \citep{Gaia_DR3:2023}.

Instrumental polarization was found to be negligible based on repeated 
observations of unpolarized standard stars. The polarization angle zero-point 
was calibrated using polarized standards from \citet{Turnshek:1990} and the 
catalog of \citet{Heiles:2000}. The final catalog contains reliable polarization 
measurements for 840 stars, of which 837 were matched with \textit{Gaia} DR3 
counterparts.

\subsection{Previous polarimetric works}

\begin{figure}
   \centering
   \includegraphics[width=.98\hsize]{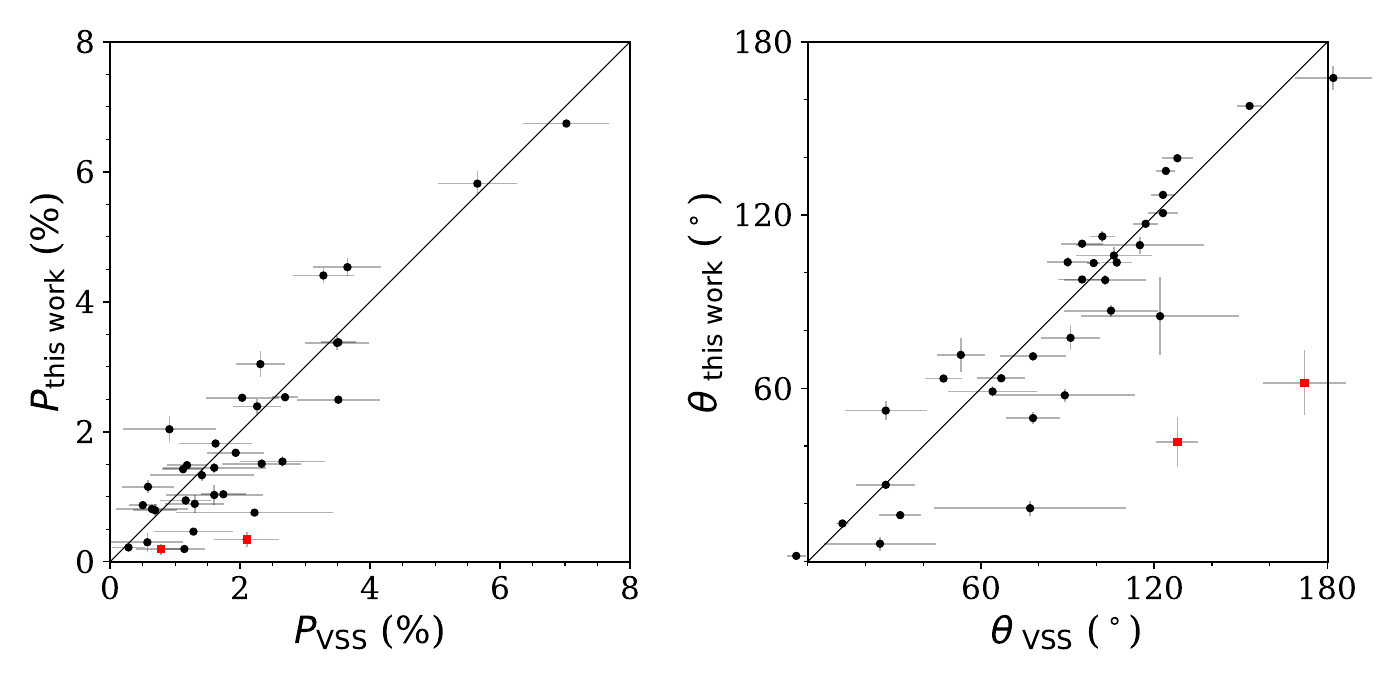}
   \caption{Comparison between the polarization measurements from \citet{Vrba:1976} and those obtained in this work 
   for the 35 stars in common. Left panel: Degree of polarization, $P$ (\%). Right panel: Polarization position angle, 
   $\theta$ (deg). The solid lines indicate the one-to-one relation and are shown for reference only. The error bars correspond 
   to the quoted uncertainties. The two discrepant data points marked by red squares are discussed in the text.}
   \label{fig:polar_vss}
\end{figure}

The magnetic field geometry for the region around NGC~1333 was previously
investigated by \citet{Vrba:1976}. They collected polarimetric measurements 
for 59 stars in an area of the sky, which is almost identical to the one covered 
by the present investigation. Among these stars, 33 are in common with our sample. 
Figure\,\ref{fig:polar_vss} gives a comparison between both samples. The left panel compares 
the degrees of polarization measured in the two works. The values obtained by \citet{Vrba:1976} 
for stars presenting small degrees 
of polarization are, in general, slightly higher than the one obtained in our 
sample. The observed difference may be a consequence of the fact that 
their measurements were performed with an unfiltered S-20 photomultiplier: 
their measurements cover a larger and bluer spectral band than ours. Despite this, there is good 
agreement between the obtained degree of polarization. 
Indeed, the right panel of Fig.\,\ref{fig:polar_vss}, which compares the determined polarization 
angles, indicates that both surveys agree. 
It must be mentioned that the error indicated in this diagram is rather underestimated 
for some objects. The expression, $\sigma_{\theta} = 28\fdg65\, \sigma_{p}/p$, 
derived by \citet{Serkowski:1974}, was indiscriminately adopted by \citet{Vrba:1976}. For consistency, we did the same. However, it is known that the applicability of this
expression should be considered only for measurements with a signal-to-noise 
greater than 6 \citep{Naghizadeh:1993}. 

The two discrepant data points marked by red squares in Fig.\,\ref{fig:polar_vss} warrant further 
comment. The point located near the edge of the diagram corresponds to a star with a very low degree of 
polarization ($0.78 \pm 0.37$\% in the sample of \citeauthor{Vrba:1976}, compared to $0.2 \pm 0.1$\% in our 
measurements), for which the determination of the polarization position angle is therefore uncertain in both 
datasets. The second object not only shows a significant discrepancy in the position angle but also exhibits the 
largest difference in the polarization degree, with values of $2.1\pm 0.5$\% in the \citeauthor{Vrba:1976} sample versus 
$0.3 \pm 0.1$\% in ours. This polarization degree appears unexpectedly low for a 
star located at a distance of about 1\,kpc, particularly given the relatively high interstellar extinction of 
$A_V \approx 2\fm3$ reported in the \texttt{StarHorse} catalog \citep{Anders:2022}.

A few years later, \citet{Turnshek:1980} collected polarization measurements for 24 stars in 
an area of almost $4\degr \times 4\degr$ and centered in NGC~1333. Their sample contains 
only two stars in common with ours, and the agreement between the measurements 
is very good.

\citet{Goodman:1990} also performed polarimetric measurements in the same region of the sky.
However, since they covered the full Perseus complex, the area they mapped is much 
larger than those surveyed in the previous work mentioned above and in ours. Their sample 
contains 83 objects; nevertheless, only nine lie in the same area covered in our work, 
and among them only two are in common with our sample. Again, both show good agreement.
Of the results obtained by \citet{Goodman:1990}, particular attention should be drawn to their discovery of 
a bimodal distribution of polarization angles in the Perseus region, attributed by them to two distinct dust 
distributions along the line of sight. The same was suggested by \citet{Vrba:1976}. This point is discussed in 
more detail in Sect.\,\ref{sec:los}.    

\citet{Alves:2011} gathered $J$-band polarimetry for 22 stars in the neighboring area of
IRAS\,4A. Their observations were conducted using the LIRIS infrared camera (in
polarimetric mode), mounted at the 4.2 m William Herschel Telescope (La Palma, 
Canary Islands, Spain). The obtained polarization map shows a magnetic field morphology
dominated by an ordered component, which is almost perpendicular to the field
obtained with submillimeter data for the dense molecular core associated with IRAS\,4A.
The authors interpret it as being due to the existence of two or more interstellar 
structures along the surveyed line-of-sight, in agreement with what is obtained from
molecular line data \citep{Ridge:2006, Pineda:2008}.

\section{Results}

\subsection{Polarization buildup along the line of sight}
\label{sec:los}

Stellar distance estimates allow us to examine how optical polarization builds
up along the line of sight toward NGC~1333. We use the homogeneous distances
from the \texttt{StarHorse} catalog to investigate the polarization degree
and position angle as functions of distance and to identify the stellar
populations, which can be used to characterize the intervening polarization
toward the Perseus molecular cloud.

\begin{figure}
   \centering
   \includegraphics[width=.95\hsize]{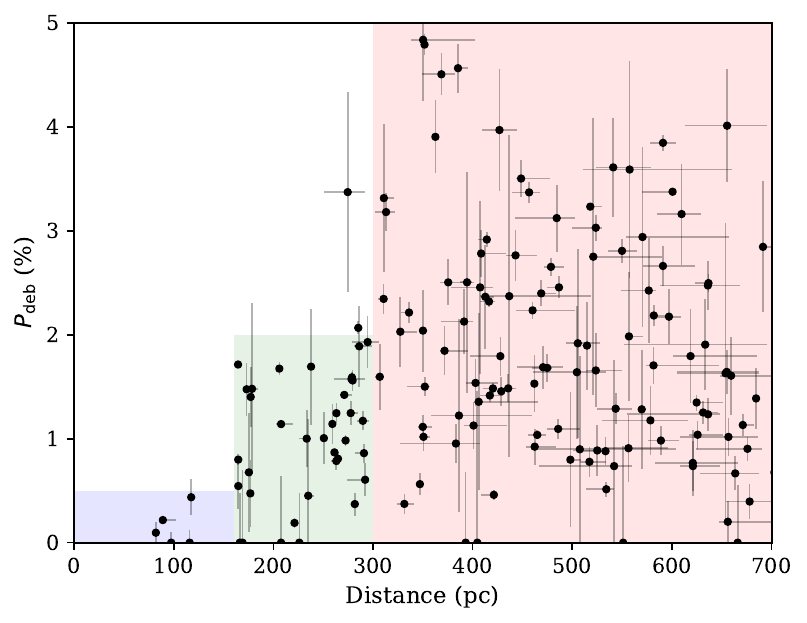}
   \caption{Debiased degree of linear starlight polarization, $P_\mathrm{deb}$, as a function of distance for
   stars projected toward the NGC~1333 region. The distances were taken from the \texttt{StarHorse}-based
   catalog \citep{Anders:2022}. The shaded areas highlight three distance intervals: a nearby population
   ($d \lesssim 160$~pc; blue), an intermediate-distance population ($\sim$160--300~pc; green), and a more distant
   population beyond 300~pc (red). The error bars indicate measurement uncertainties in polarization and distance.}
   \label{fig:polar_dist}
\end{figure}

Figure~\ref{fig:polar_dist} shows the debiased degree of linear polarization,
$P_{\rm deb}=\sqrt{P^2-\sigma_P^2}$, as a function of stellar distance.
The polarization remains relatively low for the nearest stars, with a gradual
increase becoming apparent at distances of approximately 160~pc. Between
about 160 and 300~pc, the polarization reaches intermediate values, while
stars beyond $\sim$300~pc generally show higher polarization degrees. This
behavior indicates that a substantial increase in optical polarization occurs
at distances consistent with the Perseus molecular cloud. We therefore
distinguish three broad distance regimes: a nearby, weakly polarizing
population at $d\lesssim160$~pc; an intermediate-distance population between
approximately 160 and 300~pc; and a more distant population associated with
Perseus/NGC~1333 at $d\gtrsim300$~pc. The two main distance breakpoints identified here 
are consistent with the polarization transitions at approximately 150 and 300 pc
reported by \cite{Doi:2021} across the wider Perseus region. Our more densely 
sampled optical dataset therefore provides a
detailed confirmation of this line-of-sight stratification
specifically toward NGC~1333.

\begin{figure}
   \centering
   \includegraphics[width=.95\hsize]{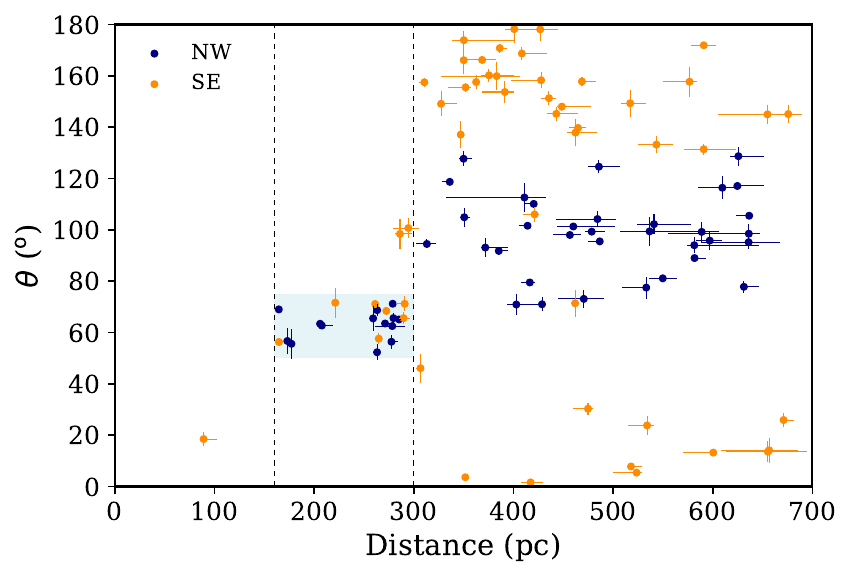}
   \caption{Polarization angle, $\theta$, as a function of distance for stars projected toward NGC~1333,
   shown separately for the NW and SE subregions. Only stars with $P/\sigma_P\geq5$ are included,
   corresponding to polarization-angle uncertainties of approximately $6^\circ$ or less. The vertical dashed
   lines mark the approximate boundaries between the nearby population ($d\lesssim160$~pc), the
   intermediate-distance population ($\sim$160--300~pc), and the more distant Perseus/NGC~1333 population
   ($d\gtrsim300$~pc).}
   \label{fig:theta_dist}
\end{figure}

The polarization position angles provide complementary information on the magnetic-field orientation 
along the line of sight. Figure~\ref{fig:theta_dist} shows their variation with distance, separately for the 
northwestern (NW) and southeastern (SE) portions of the surveyed field. To ensure reliable angle 
measurements, we retained only stars with $P/\sigma_P\geq5$, for which the polarization-angle 
uncertainties are approximately $6^\circ$ or smaller. In both regions, the angle distributions beyond 
$\sim$300~pc differ from those observed between 160 and 300~pc, suggesting that the more distant 
stellar sample probes a different dominant polarization orientation.

Two stars with nominal \texttt{StarHorse} distances of 286 and 294~pc lie close
to the adopted 300~pc boundary. Their distance uncertainties are consistent
with their association with Perseus, and their polarization angles differ
from those typical of the 160--300~pc population. We therefore exclude these
two objects from the sample used below to estimate the intervening
polarization. This exclusion is made for the foreground estimate; it does not
change the distance intervals used to display the full distance-resolved
sample in Figs.~\ref{fig:polar_dist} and \ref{fig:theta_dist}.

The contributions from different polarizing layers cannot be separated
by adding polarization degrees or position angles directly. Since the
Stokes parameters $Q$ and $U$ combine linearly, the contribution of
intervening material must be estimated in Stokes space. The polarization
degree and position angle can then be recomputed from the corrected
Stokes parameters. Thus, the observed polarization of each star reflects
the cumulative contribution of the polarizing material along the line
of sight between the observer and the star.

\begin{figure}
   \centering
   \includegraphics[width=.8\hsize]{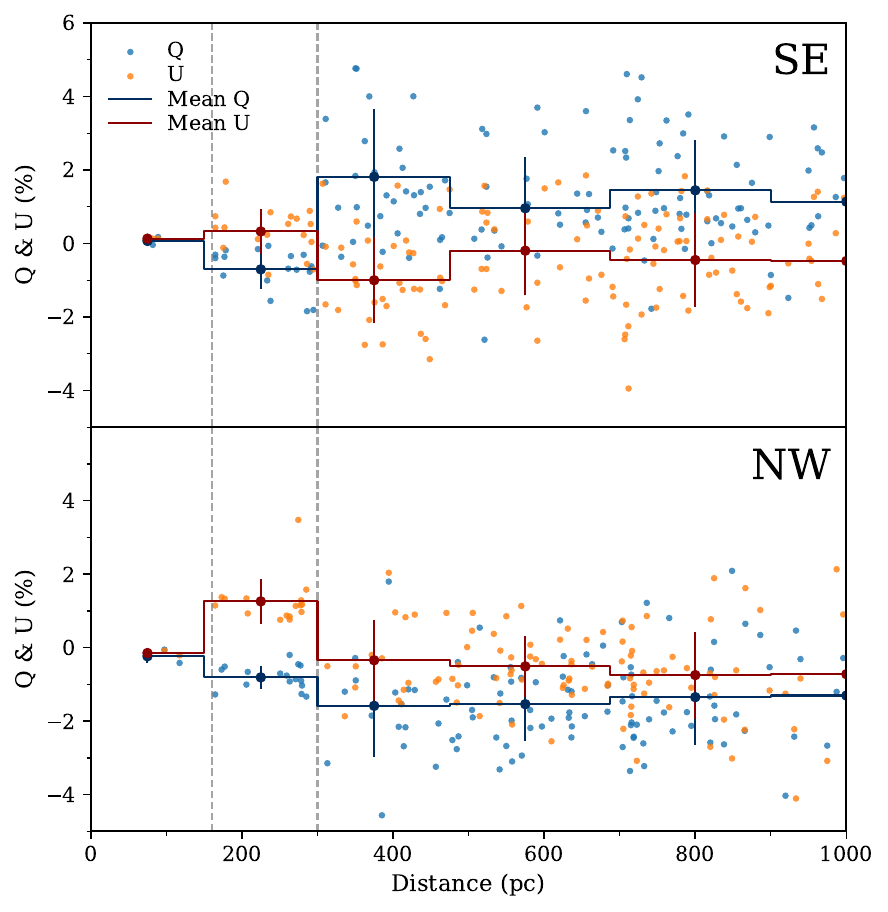}
   \caption{Stokes parameters, $Q$ and $U$, as functions of distance for the SE (upper panel) and NW 
   (lower panel) subregions. The vertical dashed lines mark distances of 160 and 300 pc. The 
   blue and orange points represent individual measurements of $Q$ and $U$, respectively, for stars 
   with $P/\sigma_P\geq1$. The step curves show the mean values of the Stokes parameters in distance 
   bins, with the error bars indicating their uncertainties. In the SE region, the binned mean $Q$ changes from 
   negative to positive values at distances of approximately 300~pc, while the binned mean $U$ changes 
   from slightly positive to negative values. These opposite sign transitions indicate a change in the dominant 
   polarization contribution along the line of sight beyond $\sim$300~pc. In the NW region, no equivalent 
   sign reversal is observed. Instead, both parameters shift toward more negative values. These contrasting 
   trends suggest spatially dependent changes in the polarization associated with the more distant stellar 
   population.}
   \label{fig:q_u}
\end{figure}

Figure~\ref{fig:q_u} displays $Q$ and $U$ as functions of distance for the NW and 
SE subregions of the surveyed field. This spatial division allows us to investigate possible 
variations in the Stokes parameters across the field. Unlike Fig.~\ref{fig:theta_dist}, this figure includes 
stars with $P/\sigma_P\geq1$. We adopt this lower threshold because the Stokes parameters remain 
useful for this analysis even when the polarization angle is relatively uncertain. The less restrictive selection 
also retains more stars at distances below 160~pc, providing a more representative view of the measurements 
across the different distance intervals. Consequently, the number of points shown in the 160--300~pc 
interval exceeds the number of stars used for the foreground estimate described below.

The scatter in $Q$ and $U$ within the intermediate-distance interval should
not, by itself, be interpreted as evidence of multiple line-of-sight
components. Stars at different distances also sample different positions on
the sky, so the scatter may reflect measurement uncertainties, spatial
variation, intrinsic dispersion, distance dependence, or a combination of
these effects. For the maximum-likelihood estimate described in
Sect.~\ref{sec:morphology}, we used the more restrictive sample of 21 stars
with $160<d<300$~pc and $P/\sigma_P\geq5$, after excluding the two objects
discussed above. This estimation sample comprises 14 stars in the NW region
and seven in the SE region. The distinction between the full sample plotted
in Fig.~\ref{fig:q_u}, and the smaller sample used for the estimate is
important: the figure visualizes the measured Stokes parameters, whereas
the high signal-to-noise subset is used to determine the regional mean
foreground Stokes vectors.

\subsection{Comparison between optical and Planck polarization}
\label{sec:optical_planck}

We next compared the observed stellar polarization, before subtracting any
estimated intervening contribution, with the polarization inferred from the
\textit{Planck} 353~GHz dust-emission data. This comparison provides a
distance-resolved diagnostic of how the optical measurements relate to the
large-scale orientation traced by \textit{Planck}. The estimate and subtraction
of the intervening Stokes contribution are presented in
Sect.~\ref{sec:morphology}.

\begin{figure}
   \centering
   \includegraphics[width=.95\hsize]{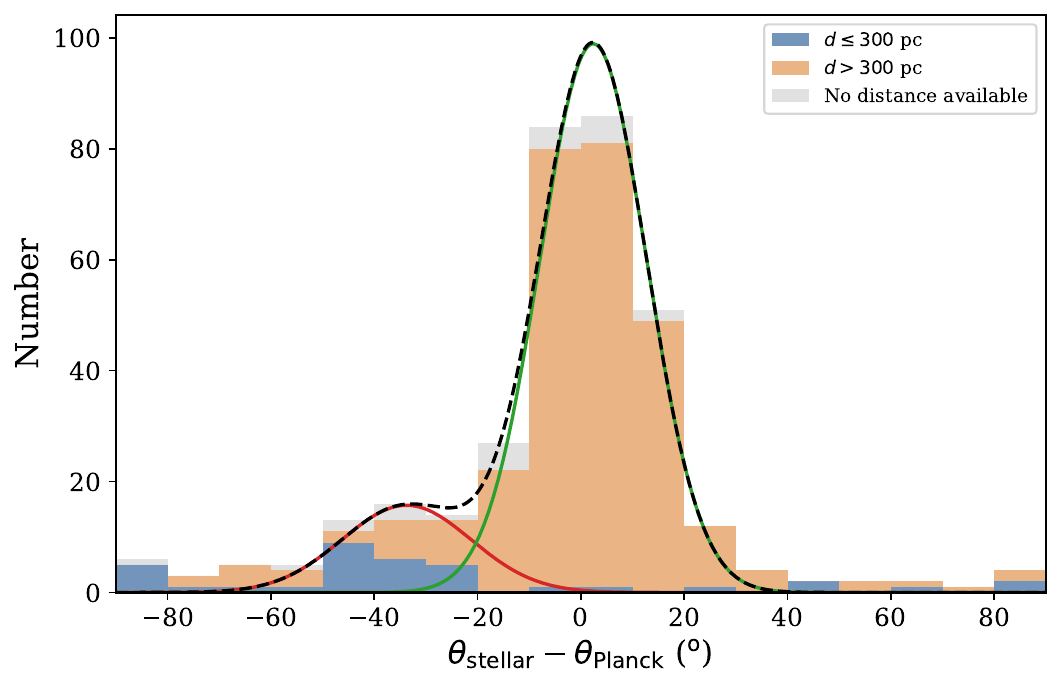}
   \caption{Distribution of angular differences $\Delta\theta=\theta_\mathrm{stellar}-\theta_\mathrm{Planck}$ between
   the observed stellar polarization and the \textit{Planck} 353~GHz polarization, using $10^\circ$ bins. The histogram
   includes stars satisfying $P/\sigma_P\geq5$, irrespective of distance. The stars are separated according to their adopted
   distance: $d\leq300$~pc (blue), $d>300$~pc (orange), and no available distance estimate from either \texttt{StarHorse} or
   \textit{Gaia} (gray). The dashed black curve shows the two-component Gaussian mixture model, with the individual
   components indicated by the colored curves. The main component near $\Delta\theta\simeq0^\circ$ is dominated by
   stars beyond 300~pc, whereas the secondary component near $\Delta\theta\simeq-40^\circ$ is predominantly contributed
   by stars at $d\leq300$~pc.}
   \label{fig:dif_theta}
\end{figure}

Figure~\ref{fig:dif_theta} shows the angular differences
$\Delta\theta=\theta_\mathrm{stellar}-\theta_\mathrm{Planck}$, where
$\theta_\mathrm{Planck}$ is the plane-of-sky (POS) magnetic-field orientation
inferred from the \textit{Planck} 353~GHz polarization after the conventional
$90^\circ$ rotation. The distribution has a prominent peak near
$\Delta\theta\simeq0^\circ$, indicating that the observed stellar
polarization angle is close to the large-scale orientation inferred from
\textit{Planck} for a substantial fraction of the sample. This component is
dominated by stars beyond 300~pc. A secondary component near
$\Delta\theta\simeq-40^\circ$ is predominantly contributed by stars at
$d\leq300$~pc, consistent with the different polarization orientations
sampled at different distances.

The comparison requires care because the two tracers do not sample the
line of sight in the same way. The polarization of each star is accumulated
only through the material between the observer and that star, whereas the
\textit{Planck} 353~GHz signal integrates polarized dust emission along the
entire line of sight, including material behind individual stars. In
addition, polarization angles are not additive: contributions from
different layers combine through their Stokes parameters. The angular
difference between a foreground-dominated stellar measurement and the
integrated \textit{Planck} orientation therefore need not equal the
position angle of any single intervening layer.

NGC~1333 lies at an intermediate Galactic latitude ($b\approx-20.5^\circ$).
The distance-resolved optical polarimetry and the extinction--distance
information available from the \texttt{StarHorse} catalog (Appendix~B)
show no clear evidence of an additional substantial polarizing structure
behind the Perseus cloud. This supports, but does not prove, the
interpretation that the large-scale \textit{Planck} orientation is dominated
by the intervening layers identified in our analysis and the
Perseus/NGC~1333 component. A weak or diffuse contribution from more distant
material cannot be excluded. We therefore regard the optical--\textit{Planck}
agreement as evidence of a consistent large-scale projected orientation,
rather than proof that the two measurements isolate identical material.

\subsection{Foreground estimate, correction, and magnetic-field morphology}
\label{sec:morphology}

We now discuss estimating the cumulative intervening polarization from the
160--300~pc stellar sample and subtract the corresponding Stokes contribution
from stars beyond 300~pc. As explained in Sect.~\ref{sec:los}, the estimate
was performed in Stokes space because the contributions from different
polarizing layers combine in $Q$ and $U$, not by direct addition or
subtraction of polarization degrees and angles. We estimated separate
foreground Stokes vectors for the NW and SE subregions, allowing for the
observed spatial difference between their mean Stokes values.

For the maximum-likelihood estimate, we used the 21 stars with
\texttt{StarHorse} distances between 160 and 300~pc, satisfying
$P/\sigma_P\geq5$, after excluding the two objects described in
Sect.~\ref{sec:los}. The NW sample comprises 14 stars and yields
\begin{equation}
Q_{\rm fg}^{\rm NW}=-0.866\pm0.088\%,
\qquad
U_{\rm fg}^{\rm NW}=1.128\pm0.062\%,
\end{equation}
corresponding to
\begin{equation}
P_{\rm fg}^{\rm NW}=1.422\pm0.072\%,
\qquad
\theta_{\rm fg}^{\rm NW}=63.76^\circ\pm1.59^\circ.
\end{equation}
For the SE region, comprising seven stars, we obtain
\begin{equation}
Q_{\rm fg}^{\rm SE}=-0.519\pm0.087\%,
\qquad
U_{\rm fg}^{\rm SE}=0.597\pm0.087\%,
\end{equation}
with
\begin{equation}
P_{\rm fg}^{\rm SE}=0.791\pm0.087\%,
\qquad
\theta_{\rm fg}^{\rm SE}=65.50^\circ\pm3.16^\circ.
\end{equation}

The mean foreground polarization degree differs significantly between the
two subregions, whereas the mean position angles are statistically
consistent. The angle difference is $1.74^\circ\pm3.54^\circ$, while the
difference in polarization degree is $0.631\pm0.113\%$. Thus, the foreground
has a similar mean orientation in the two regions but a different amplitude.
This supports estimating separate regional Stokes vectors rather than
assuming a single uniform vector across the entire survey. It does not,
however, establish that the foreground is spatially constant within either
subregion.  Each estimate represents the mean contribution of its selected
sample.

We assessed the adequacy of a constant foreground Stokes vector within each subregion 
by comparing constant models with models including linear trends with distance and sky 
position. Given the small samples (14 stars in the NW region and seven in the SE region), 
the tests do not provide robust evidence of systematic trends after accounting for multiple 
comparisons, although possible spatial dependence remains. The inferred intrinsic 
dispersions are 0.308\% and 0.209\% in $Q$ and $U$, respectively, for the NW region, 
and 0.223\% and 0.224\% for the SE region. We therefore retained a constant foreground 
vector within each subregion as a first-order approximation, while acknowledging that 
unresolved sightline-to-sightline variations may contribute additional systematic uncertainty.

To quantify the possible foreground trends, we fitted each Stokes
parameter separately as a function of distance, right ascension,
and declination, considering the NW and SE samples independently.
We used weighted linear regressions, with effective variances
$\sigma_{\rm eff}^{2}=\sigma_{\rm meas}^{2}+\sigma_{\rm int}^{2}$,
where $\sigma_{\rm int}$ is the intrinsic dispersion estimated
above. For each predictor, the linear model was compared with a
constant model. The uncertainties of the fitted slopes and the
corresponding two-sided $p$ values were obtained using the
weighted residual variance and $N-2$ degrees of freedom.
The results are summarized in Table~\ref{tab:foreground_trends}.

We applied the Holm--Bonferroni \citep{Holm:1979} correction to the 12 tests,
corresponding to two Stokes parameters, three predictors, and
two subregions. The trend of $U$ with declination in the NW
region has an uncorrected $p$ value below 0.05, while the
corresponding trend in the SE region is marginal. Neither
remains statistically significant after accounting for multiple
comparisons. We therefore retained a constant regional foreground
Stokes vector as a first-order approximation, while recognizing
that the small sample sizes do not rule out unresolved
sightline-to-sightline variations.

\begin{table*}
\caption{Linear trends in the foreground Stokes parameters.}
\label{tab:foreground_trends}
\centering
\begin{tabular}{ccccccccccc}
\hline\hline
\multicolumn{5}{c}{NW} & & \multicolumn{5}{c}{SE} \\
\cline{1-5}  \cline{7-11} 
Stokes & Predictor & Slope & $p$ & $p_{\rm H}$ & & Stokes & Predictor & Slope & $p$ & $p_{\rm H}$  \\
\cline{1-5}  \cline{7-11} 
 $Q$ & Distance & $-0.129 \pm 0.213$ & 0.558 & 1.000  &  & $Q$ & Distance & $-0.236 \pm 0.197$ & 0.285 & 1.000 \\
 $Q$ & RA       & $+0.034 \pm 0.041$ & 0.426 & 1.000 & &  $Q$ & RA       & $-0.074 \pm 0.092$ & 0.458 & 1.000 \\
 $Q$ & Dec      & $-0.093 \pm 0.079$ & 0.261 &1.000 &  & $Q$ & Dec      & $+0.008 \pm 0.035$ & 0.821 & 1.000 \\
 $U$ & Distance & $-0.140 \pm 0.153$ & 0.377 & 1.000  &  & $U$ & Distance & $-0.128 \pm 0.274$ & 0.660 & 1.000 \\
 $U$ & RA       & $+0.012 \pm 0.030$ & 0.691 & 1.000 &  & $U$ & RA       & $+0.075 \pm 0.118$ & 0.550 & 1.000 \\
 $U$ & Dec      & $+0.131 \pm 0.048$ & 0.017 &  0.204 &  & $U$ & Dec      & $+0.073 \pm 0.029$ & 0.055 & 0.607 \\
\hline
\end{tabular}
\tablefoot{The slopes are expressed in percentage points per 100 pc in
distance and per $0\fdg1$ in right ascension and declination.
The $p$ values are two-sided. The column, $p_{\rm H}$, lists the
Holm--Bonferroni adjusted values for the 12 tests.}
\end{table*}

We subtracted the corresponding regional foreground contribution directly in Stokes space:
\begin{equation}
Q_{\rm corr}=Q_{\rm obs}-Q_{\rm fg},
\qquad
U_{\rm corr}=U_{\rm obs}-U_{\rm fg}.
\end{equation}
Then, we recomputed the polarization degree and position angle from the corrected 
Stokes parameters. The uncertainties in the estimated mean foreground Stokes parameters 
were propagated through the foreground correction to the uncertainties in the corrected 
polarization degrees and polarization position angles. The correction was applied to stars 
beyond 300~pc, which were used to trace the magnetic field associated with the Perseus 
molecular cloud. This procedure estimates and removes the cumulative intervening contribution represented 
by the selected foreground stars; it does not constitute a complete 3D 
separation of all material along the line of sight. The corrected polarization is therefore 
interpreted as being associated with, and expected to be dominated by, the 
Perseus/NGC~1333 component rather than as an uncontaminated measurement of that 
component alone.

\begin{figure}
   \centering
   \includegraphics[width=\hsize]{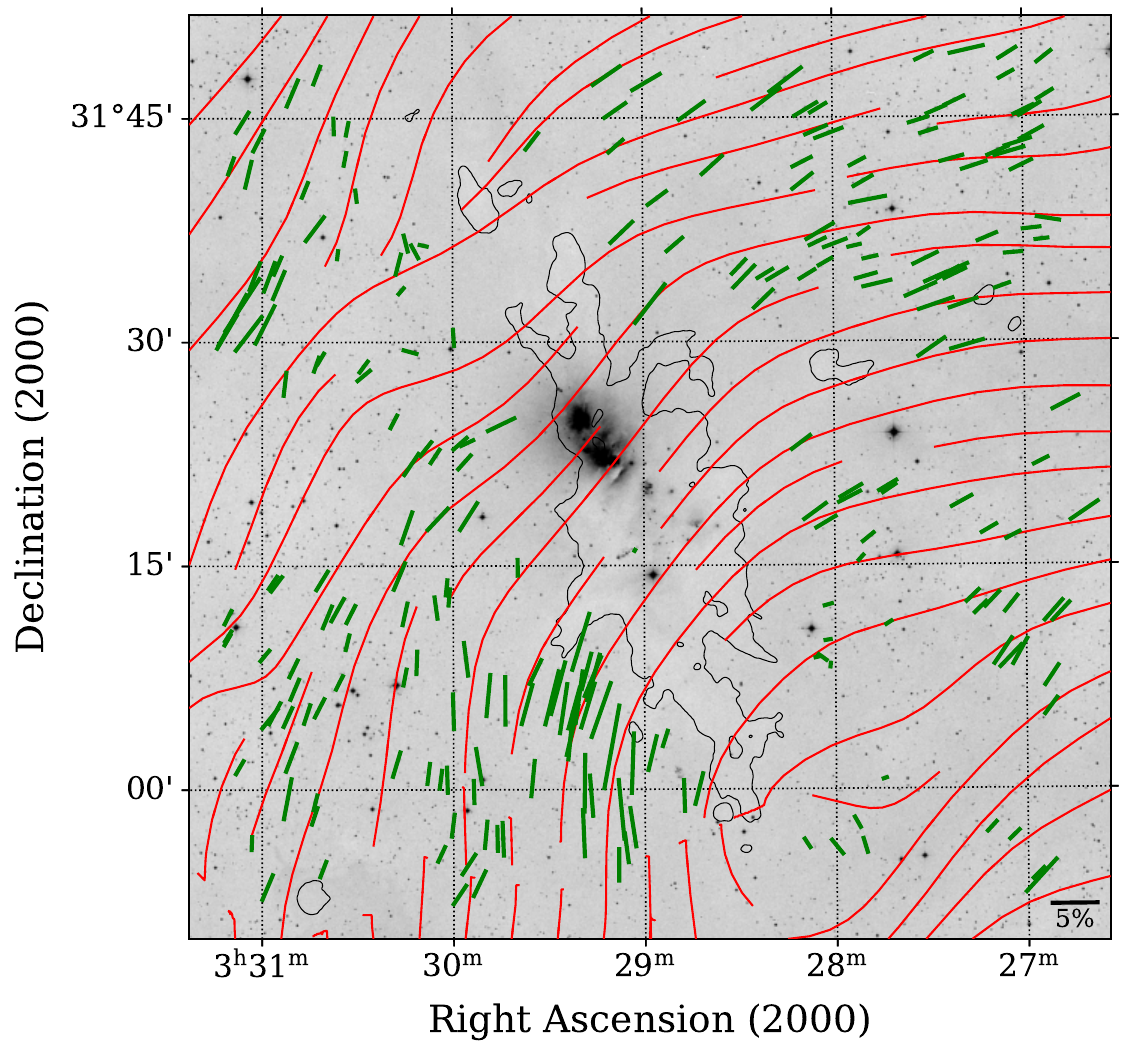}
   \caption{POS magnetic-field orientations inferred from the foreground-corrected optical stellar
   polarization measurements (green segments) overlaid on the DSS $R$-band image. Only stars
   located beyond 300~pc and satisfying $P_{\rm raw}/\sigma_{P,\rm raw}\geq5$ are shown. The red
   curves represent the magnetic-field orientations inferred from the \textit{Planck} 353~GHz dust-emission
   polarization after applying the conventional $90^\circ$ rotation. The black contours correspond to an
   H$_2$ column density of $N(\mathrm{H}_2)=5\times10^{21}\,\mathrm{cm}^{-2}$
   \citep{Andre:2010,Pezzuto:2021}. The scale of the optical polarization segments is shown in the lower-right corner.}
   \label{fig:polmap}
\end{figure}

Figure~\ref{fig:polmap} shows the foreground-corrected optical polarization
vectors of stars beyond 300~pc satisfying
$P_{\rm raw}/\sigma_{P,\rm raw}\geq5$, where ``raw'' denotes the
polarimetric quantities before foreground subtraction. The vectors are
overlaid on the DSS $R$-band image. The black contour corresponds to a
molecular hydrogen column density of
$N({\rm H}_2)=5\times10^{21}\ {\rm cm^{-2}}$
\citep{Pezzuto:2021}, a threshold commonly associated with the onset of
star formation \citep{Clark:2014}. The red curves show the POS
magnetic-field orientations inferred from the \textit{Planck} 353~GHz
dust-emission polarization after the conventional $90^\circ$ rotation.

The corrected optical vectors reveal a coherent large-scale magnetic-field
pattern across the surveyed region. The field orientation changes
systematically from the SE to the NW part of the cloud,
from a predominantly north--south orientation in the SE toward a
more east--west orientation in the NW. A similar large-scale
reorientation is visible in the \textit{Planck} map. This correspondence
supports a consistent projected field pattern on large scales. However,
because the two tracers differ in angular resolution and line-of-sight
integration, the \textit{Planck} data cannot by themselves determine whether
the reorientation is smooth or occurs over a sharper spatial transition.

To quantify this spatial variation, we defined a 1D coordinate
$s$ along the principal axis of the stellar spatial distribution,
approximately corresponding to the SE--NW direction. The systematic
variation of the optical and \textit{Planck} magnetic-field orientations
along this coordinate is quantified in Appendix~\ref{app:mf_rot}. The
binned optical orientations exhibit an approximately linear trend over
most of the surveyed region, while the first spatial bin departs from
this trend. This analysis characterizes the large-scale reorientation
without requiring a single, linear relation to describe the entire region.

For the quantitative analysis of the corrected magnetic-field orientations,
we used the NW and SE subregions introduced in Sect.~\ref{sec:los}. These
spatial subdivisions provide a practical framework for quantifying and
comparing the field properties across the mapped area; they are not
intended to imply that the field comprises two physically independent
magnetic structures.

\subsection{Quantifying the magnetic-field properties of the NW and SE regions}
\label{sec:nwse}

\begin{figure}
   \centering
   \includegraphics[width=.98\hsize]{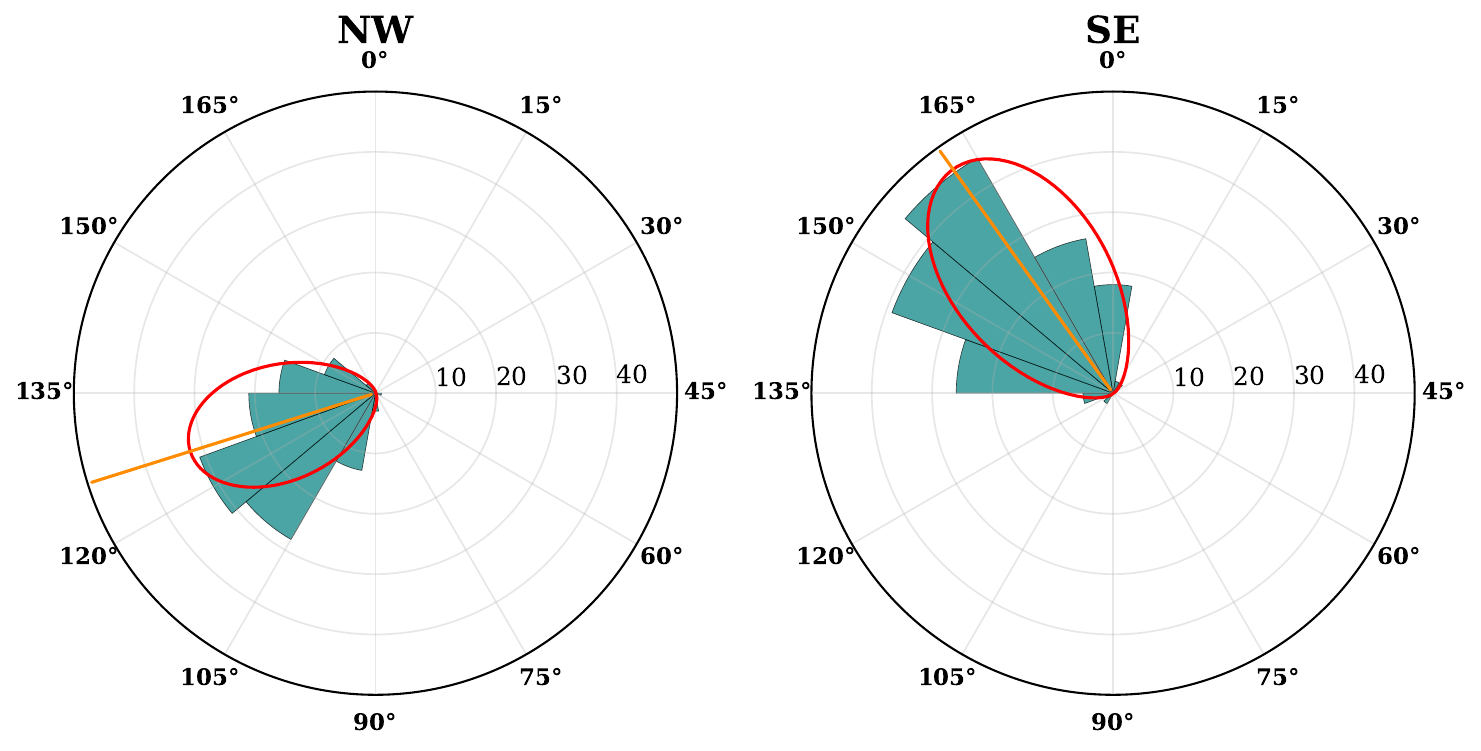}
 \caption{Polar (rose) diagrams of the foreground-corrected polarization position
   angles for the two subsamples defined in
   Sect.~\ref{sec:morphology}. Only stars located beyond 300~pc are
   included. Left panel: NW region. Right panel: SE region. The histograms
   were constructed using the axial transformation
   ($\theta \rightarrow 2\theta$) such that the angles are defined in the
   range $0^\circ$--$180^\circ$. The red curves show the best fitting von
   Mises distributions, scaled to match the observed counts. The NW
   subsample exhibits a mean direction of $\mu = 126\fdg28$, with a
   high angular concentration ($R = 0.854$, $\kappa = 3.76$). The SE
   subsample has a mean direction of $\mu = 162\fdg22$ and is similarly
   well concentrated ($R = 0.870$, $\kappa = 4.18$). The orange lines
   indicate the mean orientation in each region.
 }  
 \label{fig:rose_dia}
\end{figure}

In the following analysis, we consider only stars beyond 300 pc, with polarization measurements 
corrected for the foreground contribution estimated in Sect.~\ref{sec:morphology}.
Figure~\ref{fig:rose_dia} presents the polarization-angle distributions for the NW and SE regions. 
We used circular statistics appropriate for axial data to characterize the mean magnetic-field 
orientation, the resultant vector length ($R$), and the concentration parameter ($\kappa$) of the
corresponding von Mises distribution \citep{Mardia:2000}.

The NW region has a mean magnetic-field orientation of
$\langle\theta\rangle = 126\fdg3 \pm 0\fdg8$, while the SE region
has $\langle\theta\rangle = 162\fdg2 \pm 0\fdg6$.
The corresponding circular statistics are $R = 0.854 \pm 0.047$ and
$\kappa = 3.76 \pm 1.18$ for the NW region, and
$R = 0.870 \pm 0.038$ and $\kappa = 4.18 \pm 1.21$ for the SE region.
The similarly high values of $R$ and $\kappa$ indicate that the
polarization angles are strongly concentrated around their respective mean directions, 
despite the difference in mean orientation between the two regions.

While the polarization-angle distributions characterize the global field orientation, 
they do not quantify small-scale angular disorder around the mean field direction. 
We therefore apply the ADF analysis, following the 
methodology described in Appendix~\ref{app:method_adf}.

Figure~\ref{fig:adf} presents the ADFs obtained for the NW and SE regions together 
with their best fitting models. Both regions display the expected increase in angular 
dispersion with increasing separation, with different fitted dispersion parameters.
The NW region yields a turbulent angular dispersion parameter of
$b=0.076$\,rad ($b = 4\fdg35$), whereas the SE region presents $b=0.134$\,rad ($b = 7\fdg68$).

Using the formalism described by
Hildebrand et al.~(2009), these values correspond to
turbulent-to-ordered magnetic-field strength ratios of

\begin{equation*}
\frac{\left< B_t^2 \right>^{1/2}}{B_0} = 0.054^{+0.025}_{-0.022} \hskip1.cm \mathrm{for ~ the ~ NW ~ region}
\end{equation*}
and
\begin{equation*}
\frac{\left< B_t^2 \right>^{1/2}}{B_0} = 0.095^{+0.021}_{-0.047}  \hskip1.cm \mathrm{for ~ the ~ SE ~ region}.
\end{equation*}

Using the modified Chandrasekhar--Fermi method described in
Appendix~\ref{app:method_adf}, we obtain the POS magnetic-field
strengths of $B_{\rm POS}=59^{+44}_{-23}~\mu$G and
$B_{\rm POS}=46^{+50}_{-13}~\mu$G for the NW and SE regions, respectively.
The two values are consistent within their uncertainties.
These estimates should be regarded as approximate, since
residual sightline-to-sightline foreground fluctuations may
contribute to the measured angular dispersion and are not
fully accounted for in the quoted uncertainties.

\begin{figure}
   \centering
   \includegraphics[width=\hsize]{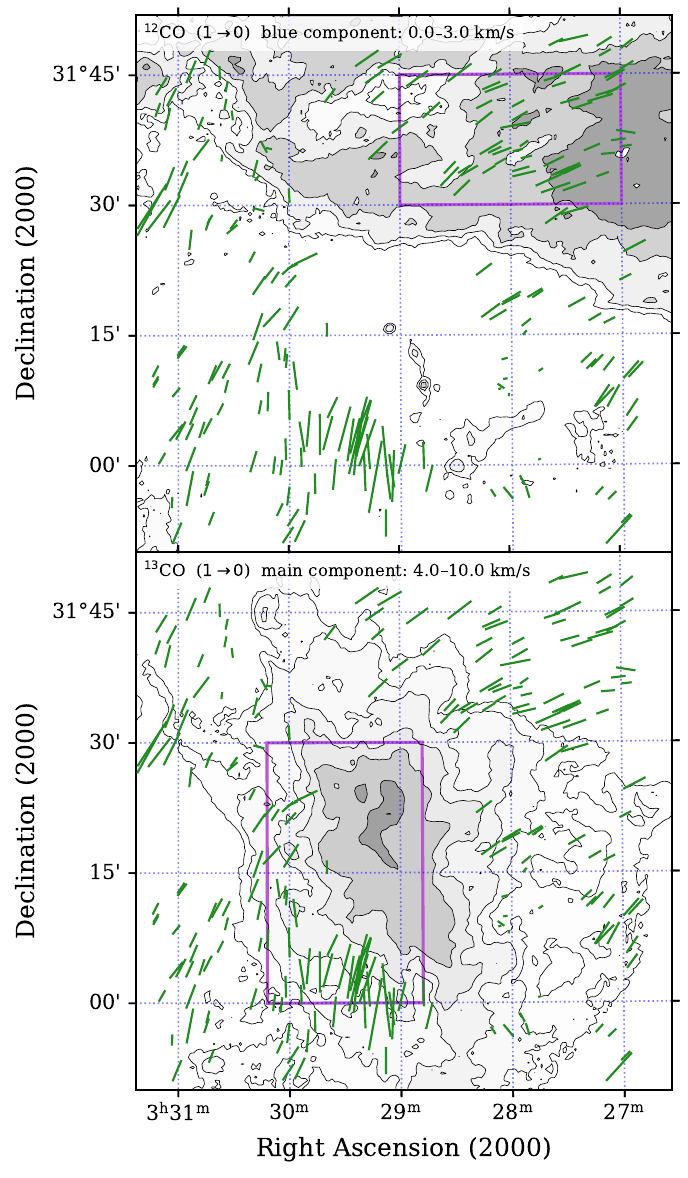}
   \caption{Foreground-corrected POS magnetic-field orientations inferred
   from optical stellar polarization measurements overlaid on the integrated CO
   emission toward the NGC~1333 region. Only stars located beyond 300~pc
   and satisfying $P_{\rm raw}/\sigma_{P,\rm raw}\geq5$ are shown.
   \textit{Top panel:} $^{12}$CO($J=1\rightarrow0$) emission integrated over the
   low-velocity interval 0.0--3.0~km\,s$^{-1}$.
   \textit{Bottom panel:} $^{13}$CO($J=1\rightarrow0$) emission integrated over the main
   velocity interval 4.0--10.0~km\,s$^{-1}$.
   The violet rectangles indicate the NW and SE regions from which the
   average CO spectra shown in Figs.~\ref{fig:co_lines} were extracted.}
   \label{fig:co12_13}
\end{figure}

\subsection{Magnetic field and molecular gas kinematics}
\label{sec:mag_fil_co}

\begin{figure}
   \centering
   \includegraphics[width=.47\hsize]{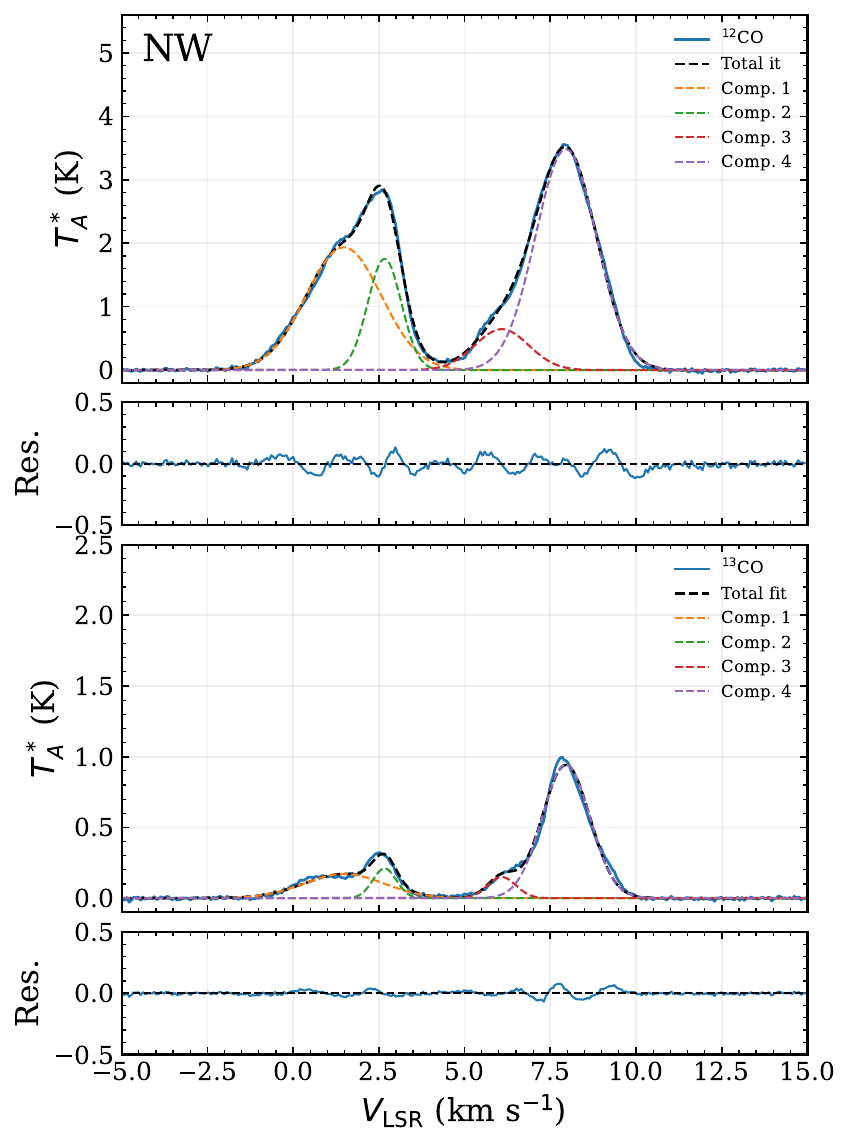}
   \includegraphics[width=.47\hsize]{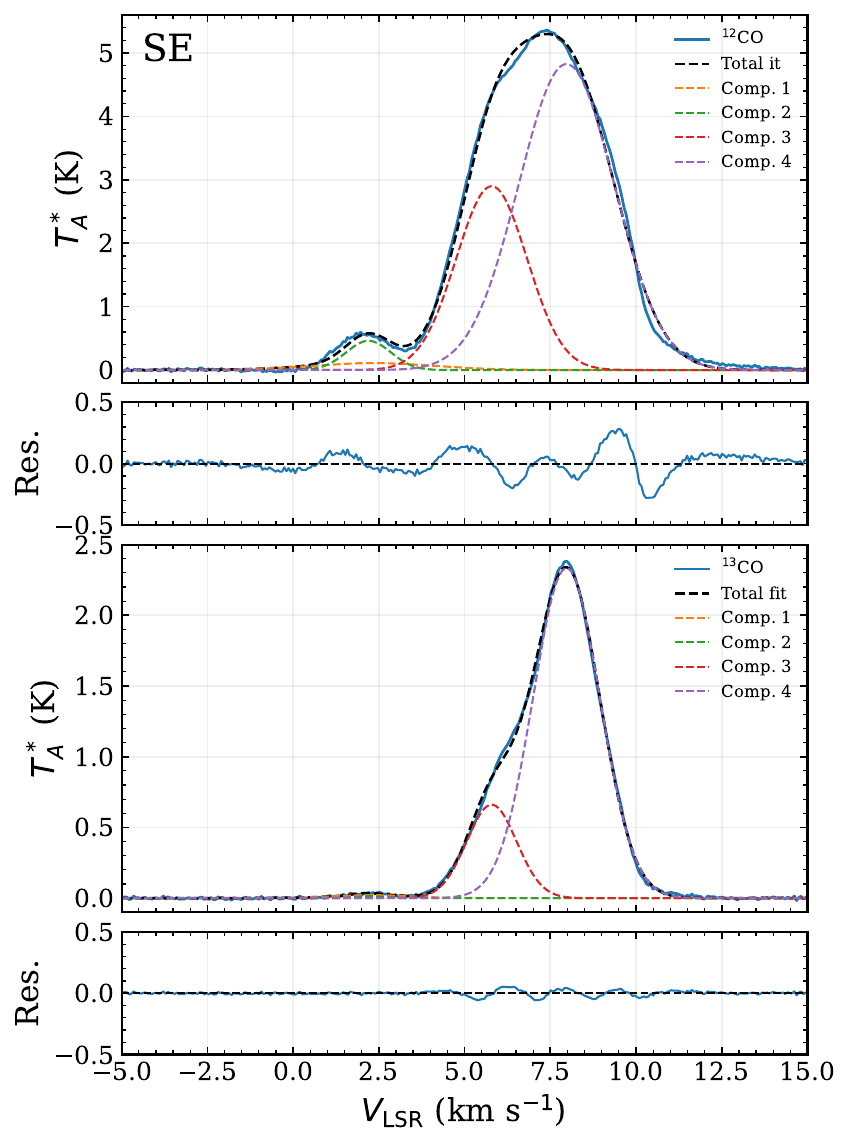}
   \caption{Multicomponent Gaussian decomposition of the $^{12}$CO and
   $^{13}$CO spectra toward the NW (left panels) and SE (right panels)
   regions. The observed spectra are shown as solid blue curves, while the dashed
   black curves represent the total best-fit models. The individual
   Gaussian components are displayed as colored dashed curves,
   corresponding to the velocity components listed in
   Table~\ref{tab:co_components}: Component 1 (orange), Component 2
   (green), Component 3 (red), and Component 4 (violet). For each region,
   the $^{12}$CO and $^{13}$CO spectra were fitted simultaneously, with
   the centroid velocity of each component constrained to be the same for
   both isotopologs, while the amplitudes and line widths were fitted
   independently. The residuals obtained after subtracting the best-fit
   models are shown in the lower panels of each spectrum.}
   \label{fig:co_lines}
\end{figure}

We next examine the molecular gas kinematics traced by the CO observations in the 
NW and SE regions. Figure~\ref{fig:co12_13} compares the foreground-corrected 
polarization segments with the molecular gas traced by the $^{12}$CO($J=1\rightarrow0$)
and $^{13}$CO($J=1\rightarrow0$) emission \citep{Ridge:2006,Pineda:2008}.
The upper panel shows the integrated $^{12}$CO emission over the
low-velocity interval $0$--$3~{\rm km\,s^{-1}}$, whereas the lower panel
shows the integrated $^{13}$CO emission over the main-velocity interval
$4$--$10~{\rm km\,s^{-1}}$. The NW and SE regions used for the spectral
analysis are indicated by the violet rectangles. The corresponding mean
spectra are shown in Fig.~\ref{fig:co_lines}, together with their
multicomponent Gaussian decomposition. The best fitting parameters are
listed in Table~\ref{tab:co_components}.

\begin{table*}
\caption{Gaussian decomposition parameters of the CO emission lines toward the NW and SE regions.}
\label{tab:co_components}
\centering
\begin{tabular}{cccccccccccccc}
\hline\hline
 & \multicolumn{6}{c}{NW} & & \multicolumn{6}{c}{SE} \\
 \cline{2-7}  \cline{9-14} \\
 &  & \multicolumn{2}{c}{$^{12}$CO} & & \multicolumn{2}{c}{$^{13}$CO} & & & \multicolumn{2}{c}{$^{12}$CO} &  & \multicolumn{2}{c}{$^{13}$CO} \\
\cline{3-4}  \cline{6-7} \cline{10-11} \cline{13-14} \\
Comp  & Vel. &Temp & FWHM & & Temp & FWHM  & & Vel. & Temp  & FWHM & & Temp  & FWHM \\
 & km/s & K & km/s & & K & km/s &  & km/s & K & km/s & & K & km/s \\
\hline
1 &  1.48 & 1.94 & 2.58 & & 0.17  & 2.74 & & 2.20 & 0.46 & 1.53 & & 0.02 & 0.90 \\
2 &  2.67 & 1.76 & 1.15 & & 0.21 & 0.81 & &  2.30 & 0.11 & 4.24 & & 0.02 & 2.68 \\
3 &  6.08 & 0.65 & 1.85 & & 0.15 & 0.97 & &  5.79 & 2.90 & 2.43 & & 0.66 & 1.71 \\
4 &  7.96 & 3.48 & 2.09 & & 0.94 & 1.65 & &  7.97 & 4.83 & 3.28 & & 2.33 & 2.30 \\
\hline
\end{tabular}
\tablefoot{For each velocity component, the centroid velocity, peak brightness temperature, and FWHM 
are listed for the $^{12}$CO and $^{13}$CO transitions. }
\end{table*}

The spectra in both regions require several velocity components and are
well described by four Gaussian components, with characteristic
centroid velocities near $1.5$, $2.5$--$2.7$, $5.8$--$6.1$, and
$8.0~{\rm km\,s^{-1}}$. The same set of components is detected in both
the NW and SE regions, although their relative contributions differ
substantially. In particular, the two lower-velocity components are
relatively stronger toward the NW, whereas the components at
$V_{\rm LSR}\sim6$ and $8~{\rm km\,s^{-1}}$ become more prominent toward
the SE. This difference is especially clear in the $^{13}$CO spectra,
where the lower optical depth provides greater contrast between the
individual velocity components. The $^{12}$CO spectra show the same
overall behavior, although with reduced contrast owing to their larger
optical depth.

The main velocity component, centered at
$V_{\rm LSR}\simeq8~{\rm km\,s^{-1}}$ in both regions, is reasonably
associated with the molecular material of NGC~1333. Its centroid is close
to the systemic velocity of IRAS~4A, which is of order $7~{\rm km\,s^{-1}}$ 
\citep[e.g.,][]{DeSimone:2022}, providing an independent reference for associating
this velocity interval with the NGC~1333 star-forming complex. The fitted
line widths of this component also show that the $V_{\rm LSR}\sim7$
--$8~{\rm km\,s^{-1}}$ emission spans a sufficiently broad velocity range
to encompass the systemic velocity of IRAS~4A.

In contrast, the emission at lower velocities, $V_{\rm LSR}\sim1$--$3~{\rm km\,s^{-1}}$,
traces additional molecular gas along the line of sight. However, the present CO data alone do not 
constrain its distance or establish whether it is physically associated with the foreground layer 
identified through stellar polarimetry or with the NGC~1333 complex. Although the distance stratification 
of the stellar polarization reveals an intermediate-distance foreground layer at $d\sim160$--$300$ pc,
a direct association between this layer and the low-velocity CO emission cannot be established from 
the available observations.

The relative contribution of the velocity components also changes
systematically across the surveyed field. The spatial variation of the
mean CO spectra, obtained using the same five spatial bins adopted for
the magnetic-field analysis, is presented in Appendix~\ref{app:co_kin}.
The spectra show that the contribution of the low-velocity emission
increases systematically along the spatial coordinate $s$, while the
relative contribution of the main-velocity component decreases. This
behavior is quantified in Appendix~\ref{app:co_kin} through the ratio
$W_{\rm low}/W_{\rm main}$ for both $^{12}$CO and $^{13}$CO. The same
spatial direction also corresponds to the gradual rotation of the
POS magnetic field quantified in Appendix~\ref{app:mf_rot}.
Together, these results indicate that the large-scale magnetic-field
reorientation occurs across a region where the relative contribution of
distinct molecular-gas velocity components also changes systematically.

\section{Discussion}
\label{sec:discussion}

\subsection{Foreground polarization and the layered magnetic structure}
\label{sec:disc_foreground}

The distance-resolved polarization measurements show that the
optical polarization toward NGC~1333 cannot be interpreted as
arising from a single interstellar component. Instead, the
polarization builds up along the line of sight, with a intermediate-distance polarization 
contribution identified between approximately 160 and 300~pc.
The relatively uniform Stokes parameters within this distance
interval support the use of a mean Stokes vector to estimate the
polarization contribution of this intermediate-distance layer.

The distance stratification recovered here is consistent with
the two main polarization breakpoints at approximately 150 and
300 pc identified by \citet{Doi:2021} across the wider Perseus
region. \citet{Doi:2021} associated the foreground component
at approximately 150 pc with the outer edge of the Taurus
molecular cloud and identified a lower-density dust cavity
between Taurus and Perseus, centered at a distance of about
240 pc and with a characteristic size of 100--160 pc. This
structure was interpreted as corresponding to an H\,\textsc{i} shell
generally associated with the Per OB2 association. The
160--300 pc interval sampled in our study overlaps this
intermediate-distance volume. Although this provides a useful
3D context for our results, the present data
do not independently constrain the spatial distribution or
physical nature of the material within this interval.

The Stokes vector measured from stars in the 160--300 pc
interval represents the cumulative polarization produced by
all material between the observer and those stars, including
any contribution from material located at distances below
approximately 160 pc. The foreground correction described
in Sect.~3.3 subtracts the corresponding mean cumulative
Stokes contribution from stars beyond 300 pc. Thus, the
correction does not isolate only the material physically
located between 160 and 300 pc, nor does it require the
assumption that the nearer component contributes
comparatively little to the background stellar polarization.
The corrected polarization should instead be understood as
an estimate of the cloud-associated contribution after
removing the cumulative foreground component represented
by the selected intermediate-distance stars.

This distinction is important for interpreting the magnetic-field
morphology inferred from stellar polarization. The observed Stokes
parameters combine contributions from different intervening
components, whose superposition can affect both the resulting
polarization angle and its angular dispersion. Consequently, part
of the apparent angular dispersion in the uncorrected sample may
reflect line-of-sight superposition rather than fluctuations
within the cloud-associated magnetic field itself. Subtracting
the estimated foreground contribution reduces this ambiguity
and provides a more appropriate basis for investigating the
large-scale magnetic-field morphology associated with NGC~1333.

The distance information is thus central to the interpretation
of the optical polarization. Without it, the distinct foreground
contribution would be difficult to identify from the integrated
polarization measurements alone. Although the correction does
not eliminate every possible line-of-sight contribution, it
provides an observationally motivated way to reduce foreground
contamination and to compare the cloud-associated field with
independent magnetic-field tracers.

\subsection{Large-scale magnetic-field structure and degree of order}
\label{sec:field_morphology}

After foreground correction, the optical polarization reveals
a systematic change in the projected magnetic-field orientation
across the surveyed region. The mean orientation changes from
$126\fdg3\pm0\fdg8$ in the NW region to $162\fdg2\pm0\fdg6$ in 
the SE region. Rather than indicating an abrupt transition between two
independent magnetic structures, the spatial distribution of
the polarization vectors favors a continuous reorientation of
the field on cloud scales.

The circular-statistics analysis supports this interpretation.
Both regions exhibit high resultant vector lengths and
concentration parameters, indicating that their polarization
angles are strongly concentrated around their respective mean
directions. The difference in mean orientation is therefore
not accompanied by a transition to a broadly disordered
configuration. The NW and SE subdivisions are best regarded
as convenient regions for quantifying the spatial variation
of the field, rather than as evidence of two physically
independent magnetic components.

The ADF provides a complementary
measure of the fluctuations around the mean field orientation.
The inferred turbulent-to-ordered magnetic-field strength
ratios are $0.054^{+0.025}_{-0.022}$ in the NW region and
$0.095^{+0.021}_{-0.047}$ in the SE region. These values
indicate that the ordered component dominates in both regions,
although the SE region exhibits a somewhat larger turbulent
contribution. The corresponding POS magnetic-field
strengths, $59^{+44}_{-23}~\mu$G and $46^{+50}_{-13}~\mu$G,
respectively, are consistent within their uncertainties.
Thus, the large-scale change in field orientation is not
accompanied by a statistically significant difference in
the estimated magnetic-field strength.

A sensitivity test in which the estimated mean foreground
Stokes vector was perturbed within its uncertainties produced
changes in the fitted ADF parameter $b$ that were small
compared with the statistical bootstrap uncertainties for the
SE region. This indicates that the uncertainty in the adopted
regional mean foreground vector alone has a limited effect
on the fitted intercept. However, this test does not quantify
residual sightline-to-sightline foreground fluctuations or
their spatial covariance. Given the limited number of
foreground stars, these contributions cannot be reliably
constrained and cannot be assumed to be negligible. Part of
the measured angular dispersion may therefore arise from
unresolved foreground structure rather than from fluctuations
intrinsic to the magnetic field associated with NGC~1333.
Consequently, the inferred turbulent-to-ordered magnetic-field
ratios and DCF field strengths should be regarded as
approximate estimates, subject to this additional
foreground-related systematic uncertainty.

The comparison with the \textit{Planck} 353~GHz dust-polarization
data provides an independent constraint on the large-scale
morphology. Despite their different line-of-sight sampling,
the foreground-corrected optical polarization and the
\textit{Planck} magnetic-field orientation exhibit a similar
overall spatial pattern. This agreement supports the
interpretation that the optical measurements trace a
large-scale magnetic-field configuration associated with
the Perseus molecular cloud. However, because the
\textit{Planck} signal integrates polarized dust emission
along the line of sight, the correspondence in orientation
does not by itself determine the relative contribution of
each intervening component.

Taken together, the polarization morphology, circular
statistics, and ADF analysis favor a picture in which the
cloud-scale magnetic field remains predominantly ordered
while changing orientation across the surveyed region.
The physical mechanism responsible for this reorientation
cannot be uniquely established from the present data.
Nevertheless, the observed continuity of the field provides
a useful framework for examining its spatial correspondence
with the molecular-gas structure and for placing localized
magnetic-field variations in a broader context.

\subsection{Magnetic field and molecular-gas kinematics}
\label{sec:mf_co}

The CO observations provide an additional perspective on
the spatial organization of the NGC~1333 complex. The spectra
exhibit the same set of velocity components in the NW and SE
regions, but their relative contributions vary systematically
across the surveyed field. In particular, the contribution
of the lower-velocity emission increases toward one end
of the spatial axis, while the relative contribution of
the main-velocity component decreases. This behavior
indicates a gradual change in the molecular-gas structure
rather than the appearance of two clearly distinct
kinematic systems.

The spatial variation of the CO components occurs along the same 
general direction as the change in the projected POS magnetic-field orientation. 
This correspondence establishes an observational link between the spatial distribution 
of the molecular-gas velocity components and the projected magnetic-field morphology.

The physical interpretation of this relationship, however, requires caution. The emission 
at 1$-$3 $\mathrm{km\,s^{-1}}$ may arise from additional molecular material along the line 
of sight, but the present CO data do not independently constrain its distance or establish 
whether it belongs to the foreground layer identified through stellar polarimetry or to the NGC~1333 
complex itself. We therefore refrain from using the low-velocity emission as independent evidence for 
either association. Instead, we retain the observational result that the relative contributions of 
the CO velocity components and the projected magnetic-field orientation vary systematically 
across the same region. Establishing their 3D relationship will require independent 
distance constraints on the molecular gas.

Likewise, the observed spatial correspondence between the molecular-gas kinematics and the 
projected magnetic-field orientation does not uniquely identify the physical mechanism 
responsible for the field reorientation. Although this correspondence may reflect the large-scale 
organization of the gas and magnetic field, distinguishing among possible mechanisms requires 
additional constraints on the 3D gas motions and magnetic-field structure. The 
present observations therefore establish a spatial correspondence between these quantities 
but do not by themselves determine their physical relationship.

\subsection{A multi-scale view of the magnetic field}
\label{sec:disc_multiscale}

\begin{figure}
   \centering
   \includegraphics[width=\hsize]{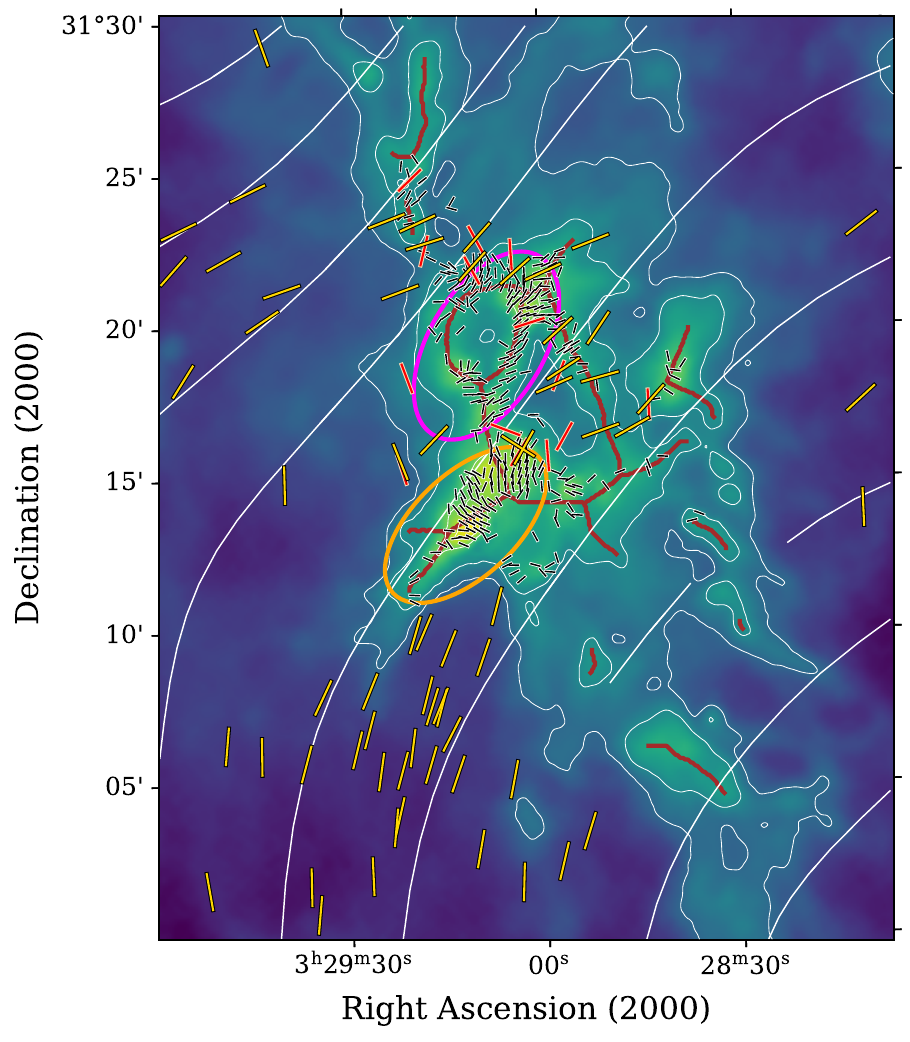}
   \caption{Comparison of the POS magnetic-field orientations
   derived from four independent polarimetric tracers toward NGC~1333.
   The foreground-corrected optical stellar polarization segments
   (yellow with black outlines) are shown only for stars with
   $d>300$~pc and $P_{\rm raw}/\sigma_{P,\rm raw}\geq5$.
   The near-infrared polarization measurements from
   \citet{Tamura:1988} are shown as red segments with orange outlines.
   Magnetic-field orientations inferred from the \textit{Planck}
   353~GHz dust polarization are represented by the gray streamlines,
   while the black segments with white outlines correspond to the
   JCMT/SCUBA-2 POL-2 dust-polarization measurements obtained by
   \citet{Doi:2020}. The white contours trace the
   $N({\rm H}_2)$ column density derived by \citet{Pezzuto:2021},
   and the red curves show selected filament skeletons derived
   in this work from the $N({\rm H}_2)$ column-density map.
   The ellipses highlight two regions discussed in
   Sect.~\ref{sec:disc_multiscale}. The comparison illustrates
   the continuity of the cloud-scale field together with
   localized differences between the magnetic field traced
   in lower-density material and that resolved within dense
   filamentary structures.}
   \label{fig:hgbs}
\end{figure}

The comparison of the four polarimetric tracers shown in
Fig.~\ref{fig:hgbs} provides complementary information on
the magnetic-field structure across different spatial scales,
gas densities, and effective line-of-sight depths. The
foreground-corrected optical polarization samples the
magnetic-field orientation in relatively diffuse material
on cloud scales. The near-infrared measurements extend
the polarization analysis toward regions of higher
extinction, while the \textit{Planck} 353~GHz polarization
traces the line-of-sight-integrated magnetic field through
thermal dust emission. The JCMT/POL-2 observations, in
turn, resolve the field within dense molecular structures
and the filamentary network associated with active
star formation.

These tracers should not be interpreted as measurements
of a single magnetic component observed at different
wavelengths. Instead, they provide complementary constraints
on the field in different physical regimes. The large-scale
orientation inferred from the foreground-corrected optical
polarization follows the overall pattern traced by
\textit{Planck} and is broadly consistent with the
near-infrared measurements, whereas the POL-2 observations
show more pronounced localized variations within dense
filamentary structures. Such differences are consistent
with previous JCMT/BISTRO observations of NGC~1333,
which revealed substantial changes in magnetic-field
morphology below cloud scales and different orientations
between individual filaments and their surrounding field
\citep{Doi:2020}. At the filament scale, \citet{Doi:2021a}
also find that the internal structure of some quiescent
filaments is consistent with a pinched magnetic field,
with field lines displaced toward the filament axis as
the material contracts.

The wide-field optical observations provide an important
constraint by extending the magnetic-field information
beyond the dense structures resolved by POL-2. They place
the localized submillimeter morphologies within the
broader field pattern, which changes systematically from
NW to SE across several parsecs. Combined
with the molecular-gas analysis in
Sect.~\ref{sec:mag_fil_co}, this provides a link between
the field orientation in the lower-density surroundings
and the spatially varying molecular structure of NGC~1333.
The optical and submillimeter measurements therefore
complement one another rather than duplicating the same
information.

The regions highlighted in Fig.~\ref{fig:hgbs} illustrate
the diversity of magnetic-field configurations at smaller
scales. Around the IRAS~4A complex (orange ellipse), the
POL-2 field is predominantly perpendicular to the
filamentary structure, whereas the optical polarization
in the surrounding lower-density material has a
substantially different orientation. The field measured
outside the dense structure therefore need not reproduce
the orientation observed within the filament itself.
This behavior is consistent with the multiscale complexity
reported by \citet{Doi:2020}, while the present optical
data show that the local configuration is embedded in
a larger-scale field whose orientation varies smoothly
across the surrounding material.

The second highlighted region (magenta ellipse) exhibits
a different configuration. The POL-2 orientations span
a broader range of angles relative to the filament
skeleton, while the optical polarization is more closely
aligned with the orientation traced in the denser material.
The relationship between the projected POS 
magnetic-field orientations in the diffuse and dense material is therefore 
not spatially uniform. In some locations, the projected field orientation 
surrounding a dense structure differs substantially from that inferred within it; 
elsewhere, their projected orientations are more closely aligned.
These comparisons necessarily concern projected
POS orientations. \citet{Doi:2020} show that 
part of the observed diversity in the relative orientations
between magnetic fields and filaments in NGC~1333 can arise
from projection effects associated with different
3D inclinations. Although their analysis
addresses the field--filament geometry rather than the
optical--POL-2 comparison made here, it illustrates an
important degeneracy: differences between the projected
optical and POL-2 position angles do not necessarily imply
an equivalent change in the 3D magnetic-field
direction between diffuse and dense material.

The present observations do not establish an evolutionary
sequence in which dense structures necessarily inherit
a preexisting cloud-scale field and subsequently develop
local departures. Instead, they show that localized
magnetic-field variations coexist with a broader,
organized magnetic structure. The optical polarization
is particularly valuable because it extends the field
information beyond the dense filaments and connects
their local submillimeter morphology to the several-parsec
organization inferred from the wide-field stellar survey.

Overall, the multi-scale comparison indicates that the
magnetic field toward NGC~1333 is organized on cloud
scales but exhibits localized changes in orientation
within the dense filamentary regime. Together with the
distance-resolved optical polarization and the molecular-gas
kinematics, these results show that the observed complexity
reflects both the superposition of polarized components
along the line of sight and  spatial variations in the projected POS
magnetic-field orientation. By estimating the
contribution of an intermediate-distance foreground
layer, the optical survey provides a clearer view of
the cloud-associated field and places the localized
structures traced by submillimeter polarization within
their broader magnetic environment.

\section{Conclusions}
\label{sec:conclusions}

We presented a wide-field optical polarimetric survey covering nearly one square degree toward the
NGC~1333 star-forming region, combining linear polarization measurements for about 840 stars with
\textit{Gaia}/\texttt{StarHorse} distances, molecular-line observations, and complementary infrared and submillimeter
polarization data. Our main conclusions are as follows:

\begin{enumerate}

\item The polarization--distance relation reveals multiple interstellar layers along the line of sight toward NGC~1333. A somewhat weakly polarizing, nearby component is followed by an intermediate-distance polarized contribution traced 
by stars at approximately 160 and 300~pc, while the main increase in polarization occurs beyond $\sim$300~pc, 
where the Perseus molecular cloud dominates the observed signal. The relatively uniform Stokes parameters 
within the intermediate-distance layer allow its polarization contribution to be estimated and subtracted in Stokes 
space. This correction reduces the contribution of this layer to the observed polarization and provides a more 
reliable basis for characterizing the magnetic-field orientation associated with the Perseus cloud. However, it does 
not fully separate all possible line-of-sight components.

\item After correcting for the estimated foreground contribution, the optical polarization reveals a coherent cloud-scale
magnetic-field morphology, exhibiting a continuous shift in orientation from the NW to the SE part of
the surveyed region. The angular-dispersion analysis indicates predominantly ordered fields in both regions but with
distinct scale-dependent behaviors: in the NW region, the ADF follows an approximately linear trend with $\ell^2$
up to angular separations of about $5'$, whereas the SE region departs from this trend at separations of about $4'$.
These differences show that the two regions are not identically described by the same angular-dispersion behavior,
despite their comparable, estimated POS magnetic-field strengths of
$59^{+44}_{-23}~\mu{\rm G}$ and $46^{+50}_{-13}~\mu{\rm G}$ in the NW and SE regions, respectively.
The smooth change in mean orientation is therefore consistent with a reorientation of the large-scale field rather
than the need for two independent magnetic structures.

\item The CO observations reveal the same molecular velocity components in the NW and SE regions but with
systematically different relative contributions. These differences accompany the large-scale change in magnetic-field
orientation and indicate that the molecular-gas distribution varies across the surveyed region. The correspondence
between the magnetic-field morphology and the molecular kinematics is consistent with a relationship between the
two, although the present observations do not establish a causal connection.

\item Our comparison with complementary polarimetric tracers shows that the foreground-corrected optical polarization
agrees well with the large-scale magnetic-field orientation traced by Planck and is broadly consistent with near-infrared
measurements. The JCMT/POL-2 observations reveal localized departures in the densest filaments and star-forming
structures. Together, these observations indicate that a coherent large-scale magnetic-field pattern coexists with local
variations in the dense gas. The comparison establishes a connection between the field orientations sampled at different
spatial scales but does not by itself determine the physical origin or evolutionary sequence of the localized departures.

\end{enumerate}

Taken together, these results show that the magnetic-field morphology toward NGC~1333 reflects both line-of-sight
superposition and spatial variations in the projected POS magnetic-field orientation across spatial scales.
By combining wide-field optical polarimetry with stellar distance information, this work provides a 
means for separating the contribution of an intermediate-distance polarized layer from the polarization 
associated with the Perseus cloud, thereby yielding a more reliable view of its large-scale 
magnetic-field structure.

\begin{acknowledgements}
We thank the anonymous referee for a careful review of the manuscript.
We are grateful to the staff of the Observatório do Pico dos Dias for their support during the observations. 
J.M.G. acknowledges  financial support from grant PID2023-146675NB-I00 (MCI-AEI-FEDER, UE). This 
work was also partially supported by 
the Unidad de Excelencia María de Maeztu programme (CEX2020-001058-M).
This work made use of data from the COMPLETE Survey \citep{Ridge:2006, Pineda:2008}, the  \textit{Gaia} mission of the 
European Space Agency (ESA; \url{https://www.cosmos.esa.int/gaia}), processed by the  \textit{Gaia} Data Processing and 
Analysis Consortium (DPAC; \url{https://www.cosmos.esa.int/web/gaia/dpac/consortium}), the  \textit{Planck} Legacy Archive 
(\url{https://pla.esac.esa.int}), an ESA science mission with instruments and contributions directly funded by ESA Member 
States, NASA, and Canada, and the Herschel Gould Belt Survey (HGBS; \url{http://gouldbelt-herschel.cea.fr}).
The HGBS is a Herschel Key Programme jointly carried out by SPIRE Specialist Astronomy Group 3 (SAG 3), scientists 
from several institutes within the PACS Consortium (CEA Saclay, INAF-IFSI Rome, INAF-Arcetri, KU Leuven, and MPIA 
Heidelberg), and scientists from the Herschel Science Center (HSC).
\end{acknowledgements}

\bibliographystyle{aa}

\bibliography{ngc1333_refs}

\begin{appendix}

\section{Spatial variation of the magnetic field and molecular-gas kinematics}
\label{app:spatial_variation}

\begin{figure}[h]
   \centering
   \includegraphics[width=.7\hsize]{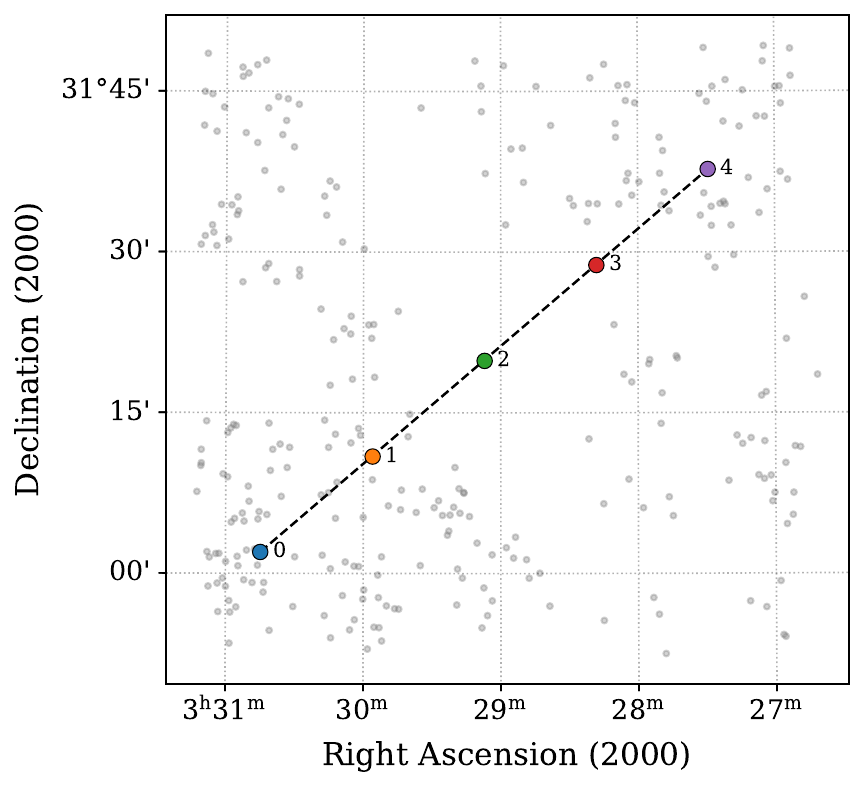}
   \caption{Spatial distribution of the selected foreground-corrected
   stellar polarization measurements and location of the five positions
   adopted for the spatially resolved analysis. The positions lie along a
   straight SE--NW axis and are equally spaced between the SE
   and NW endpoints. The numbers identify the five positions
   used throughout the appendix.}
   \label{fig:app_centers}
\end{figure}

We compare the magnetic-field orientation and CO kinematics at five equally
spaced positions along a SE--NW axis across NGC~1333. The axis
endpoints were defined from foreground-corrected stellar polarization
measurements with $d>300$~pc and $P_{\rm raw}/\sigma_{P,\rm raw}\geq5$. The positions lie
approximately 0, 825, 1649, 2474, and 3298 arcsec from the SE
endpoint (Fig.~\ref{fig:app_centers}). 

\subsection{Quantification of the spatial rotation of the magnetic field}
\label{app:mf_rot}

The spatial variation of the POS magnetic-field orientation was
determined independently from the optical stellar polarization
measurements and from the \textit{Planck} 353 GHz polarization data.

For the optical measurements, at each of the five positions we selected
the foreground-corrected stars satisfying $d>300$ pc and
$P_{\rm raw}/\sigma_{P,\rm raw}\geq5$ within a circular aperture of radius $400''$ centered
on that position. The mean magnetic-field orientation was calculated
using axial circular statistics, since polarization orientations are
defined modulo $180^\circ$. The uncertainty of each optical mean was
estimated by bootstrap resampling of the stellar sample within the
corresponding aperture.

The use of a fixed $400''$ aperture provides a spatially localized
measurement around each of the five positions while retaining a
sufficient number of stellar polarization measurements for the statistical
calculation. At the third position along the SE--NW axis, no selected
stellar polarization measurements are found within the $400''$ aperture.
Consequently, no optical mean orientation is assigned to this position.

At each position, the molecular emission is measured within a fixed $10 \times 10$-pixel 
box (approximately  $230''\times230''$) in the CO data. The \textit{Planck} polarization 
measurement is obtained from the \textit{Planck} pixels covering the 
same sky area as the corresponding CO box.

For each selected \textit{Planck} pixel, the POS magnetic-field
orientation is obtained from the Stokes $Q$ and $U$ parameters after
applying the appropriate coordinate transformation and the $90^\circ$
rotation between the polarization orientation and the magnetic-field
orientation. The mean orientation within each CO-box footprint is then
calculated using axial circular statistics. Its uncertainty is estimated
from the distribution of the valid \textit{Planck} measurements
contributing to the corresponding footprint. When more than one valid
pixel contributes, the uncertainty is obtained from bootstrap
resampling of the individual pixel orientations; for a single valid
pixel, the uncertainty is obtained from the propagated uncertainty of
the corresponding Stokes parameters.

\begin{figure}
   \centering
   \includegraphics[width=.97\hsize]{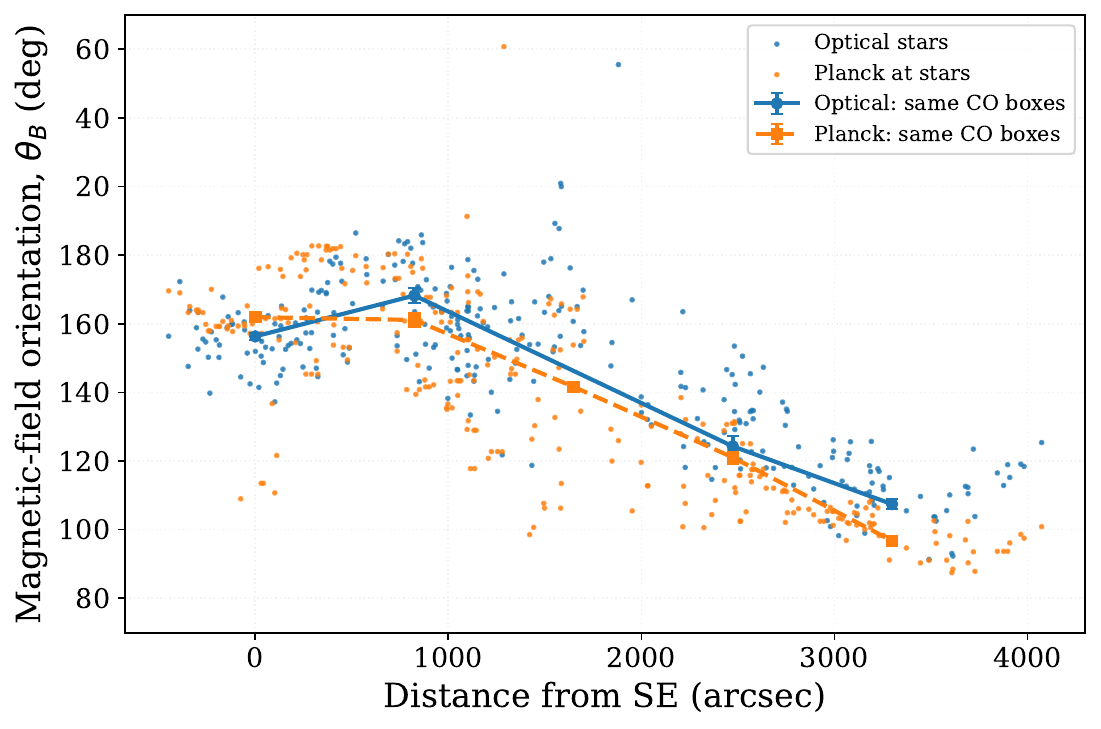}
   \caption{Spatial variation of the POS magnetic-field
   orientation across NGC~1333. The blue points show the mean optical
   magnetic-field orientations calculated from foreground-corrected stars
   satisfying $d>300$ pc and $P_{\rm raw}/\sigma_{P,\rm raw}\geq5$ within $400''$ of each of
   the five positions shown in Fig.~\ref{fig:app_centers}. The orange
   points show the corresponding \textit{Planck} 353 GHz magnetic-field
   orientations calculated from the \textit{Planck} pixels covering the
   same sky footprint as the $10\times10$-pixel CO box at each position.
   The lines connect the corresponding spatial measurements, and the
   error bars indicate their uncertainties. The horizontal coordinate is
   the distance from the SE endpoint along the SE--NW axis.
   No optical point is shown at the third position because no selected
   stellar polarization measurements are found within the $400''$
   aperture. For visualization only, orientations below $70^\circ$ are
   shifted by $180^\circ$, preserving the axial nature of the
   polarization angles.}
   \label{fig:app_rot}
\end{figure}

The resulting spatial variation of the optical and \textit{Planck}
magnetic-field orientations is shown in Fig.~\ref{fig:app_rot}. The
horizontal coordinate is the distance from the SE endpoint
along the adopted SE--NW axis. The five points therefore correspond
directly to the five spatial positions shown in Fig.~\ref{fig:app_centers}.

The five positions reveal a systematic change in the magnetic-field
orientation from the SE to the NW end of the region.
The optical and \textit{Planck} measurements show broadly similar spatial
behavior, although the two tracers do not coincide exactly at every
position. The absence of an optical measurement at the third position is
a consequence of the stellar sampling within the adopted aperture,
whereas the corresponding \textit{Planck} measurement remains available
because it is determined independently of the optical stellar sample.

For visualization only, optical and \textit{Planck} angles smaller than
$70^\circ$ are shifted by $180^\circ$ in Fig.~\ref{fig:app_rot}. This
transformation does not affect the angular statistics, uncertainties, or
angular differences used in the analysis, and is applied solely to
display the axial orientations continuously over the plotted angular
range.

\subsection{Spatial variation of the molecular-gas velocity components}
\label{app:co_kin}

We investigated the spatial variation of the molecular-gas kinematics at
the same five positions shown in Fig.~\ref{fig:app_centers}. At each
position, a $10\times10$-pixel box was extracted from the CO data cube.
The same spatial boxes are used for both $^{12}$CO and $^{13}$CO.

For each box, the molecular-line spectrum was calculated by averaging the
CO emission over the pixels contained within the box. The uncertainty of
the spatially averaged spectrum was estimated using a block-bootstrap
procedure. In this procedure, neighboring pixels were resampled in
spatial blocks rather than individually, thereby approximately accounting
for the spatial correlation introduced by the telescope beam.

\begin{figure}
    \centering
    \includegraphics[width=.97\hsize]{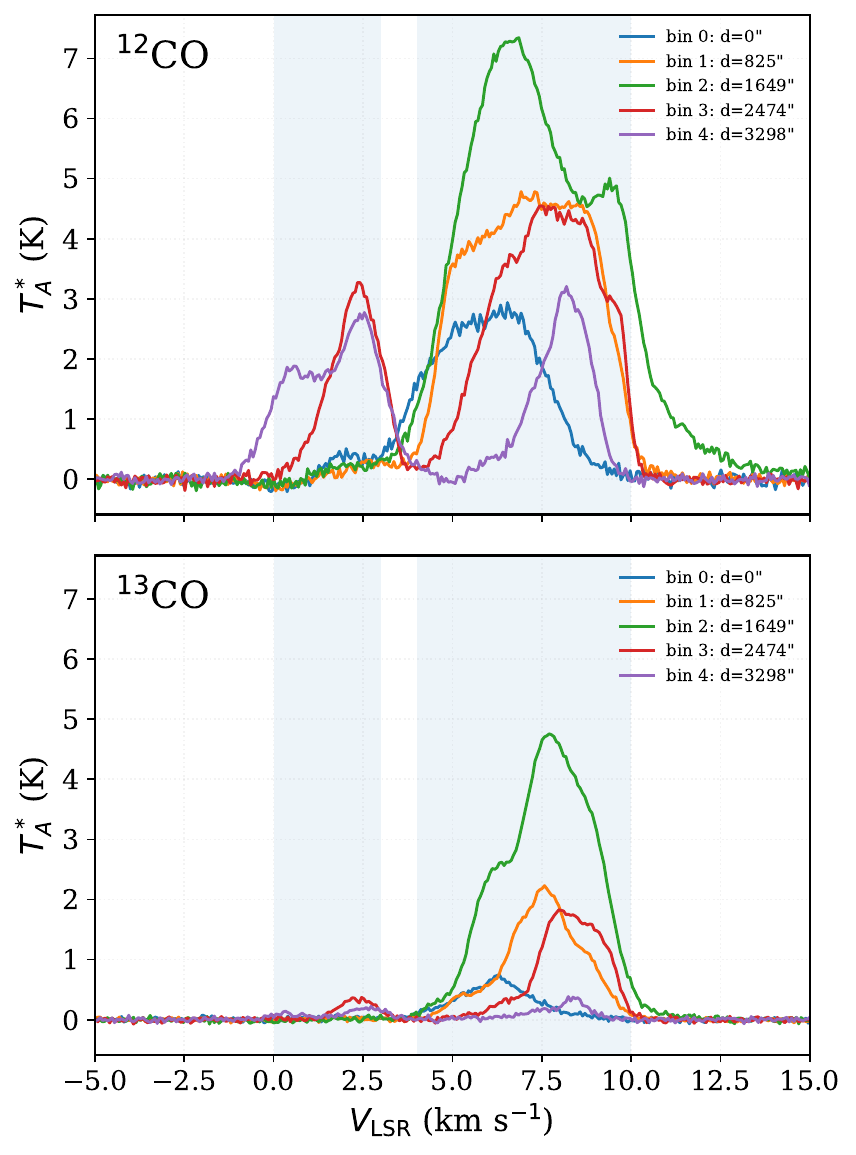}
    \caption{Mean $^{12}$CO (upper panel) and $^{13}$CO (lower panel)
    spectra extracted from the five $10\times10$-pixel CO boxes located
    along the SE--NW axis shown in Fig.~\ref{fig:app_centers}. The boxes
    are centered at distances of approximately 0, 825, 1649, 2474, and
    3298 arcsec from the SE endpoint. The shaded regions mark
    the low-velocity and main-velocity intervals adopted in the
    spatially resolved analysis.}
    \label{fig:app_co_spectra}
\end{figure}

The resulting spectra are shown in Fig.~\ref{fig:app_co_spectra}. The
five spectra correspond to the five fixed positions along the SE--NW
axis. The spectra show a spatial evolution in the relative contribution of the
different velocity components. The low-velocity emission becomes
progressively more important toward one end of the SE--NW axis, while
the component around $V_{\rm LSR}\sim7$--$8~{\rm km\,s^{-1}}$ remains
prominent across the region. This behavior is present in both
$^{12}$CO and $^{13}$CO, although the relative strengths of the
different components differ between the two isotopologs.

For the spatially resolved analysis, we define the low-velocity component
over $0$--$3~{\rm km\,s^{-1}}$ and the main component over
$4$--$10~{\rm km\,s^{-1}}$. For each spatial position, the
velocity-integrated intensities are calculated as

\begin{equation}
W_{\rm low} =
\int_{0}^{3~{\rm km\,s^{-1}}}
T_A^*(v)\,dv,
\end{equation}
and

\begin{equation}
W_{\rm main} =
\int_{4}^{10~{\rm km\,s^{-1}}}
T_A^*(v)\,dv.
\end{equation}

We then define the relative contribution of the low-velocity component as

\begin{equation}
R_{\rm CO} =
\frac{W_{\rm low}}{W_{\rm main}}.
\end{equation}

The uncertainties in $R_{\rm CO}$ are derived from the bootstrap
realizations of the spatially averaged spectra, thereby propagating the
uncertainty associated with the spatial sampling of the molecular-line
emission.

\begin{figure}
    \centering
    \includegraphics[width=.97\hsize]{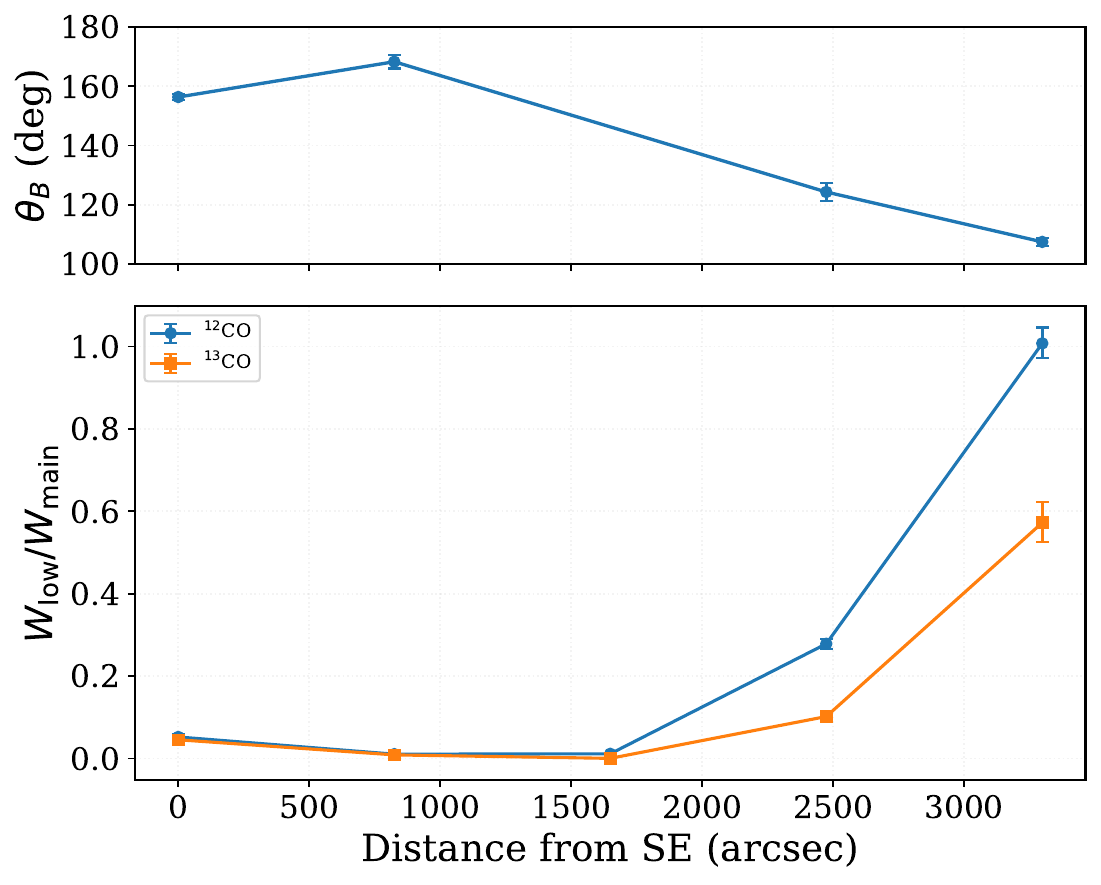}
    \caption{Spatial variation of the magnetic-field orientation and
    relative contribution of the molecular-gas velocity components along
    the SE--NW axis. Top: Optical and \textit{Planck}
    magnetic-field orientations at the five positions shown in
    Fig.~\ref{fig:app_centers}. Bottom:
    $W_{\rm low}/W_{\rm main}$ for $^{12}$CO and $^{13}$CO. Both
    quantities are evaluated at the same five spatial positions, allowing
    their spatial variations to be compared directly.}
    \label{fig:app_co_relation}
\end{figure}

The resulting spatial variation of $R_{\rm CO}$ is shown in
Fig.~\ref{fig:app_co_relation}. Both isotopologs show a systematic
change in the relative contribution of the low-velocity emission along
the SE--NW direction.

The spatial variation of $R_{\rm CO}$ occurs over the same set of
positions at which the magnetic-field orientation changes. The gradual
change in the magnetic-field orientation therefore occurs along the same
spatial direction over which the relative contribution of the
low-velocity molecular emission changes. This spatial correspondence
provides an observational basis for comparing the large-scale
magnetic-field structure with the molecular-gas velocity components.

This correspondence does not, by itself, establish a causal relation
between the gas kinematics and the magnetic-field orientation. Rather,
it shows that the relative contribution of distinct molecular-gas
velocity components and the large-scale magnetic-field orientation vary
coherently across the same region. The observed behavior is consistent
with a line-of-sight structure in which molecular material at different
velocities contributes with different relative strengths across
NGC~1333.

\section{Angular dispersion function analysis}
\label{app:method_adf}

The ADF provides a quantitative
description of the scale dependence of the polarization-angle
dispersion and allows us to characterize fluctuations of the magnetic
field around its large-scale structure. Following the formalism
introduced by \citet{Hildebrand:2009}, the ADF is defined as
\begin{equation}
\left\langle \Delta \Phi^2(\ell) \right\rangle
=
\frac{1}{N(\ell)}
\sum_{i=1}^{N(\ell)}
\left[
\Phi(\mathbf{x}_i)
-
\Phi(\mathbf{x}_i+\boldsymbol{\ell})
\right]^2,
\end{equation}
where $\Phi$ is the polarization angle and $N(\ell)$ is the number of
stellar pairs contributing to a separation bin centered on $\ell$.
Because different pairs may share one or both stars, these pair
combinations are not statistically independent. We therefore use
$N(\ell)$ only to denote the number of pair combinations in a given
separation bin, rather than the number of independent measurements.

The magnetic field can be described as the sum of a smoothly varying
large-scale component, $B_0(\mathbf{x})$, and a turbulent component,
$B_t(\mathbf{x})$, with characteristic scales $d$ and $\delta$,
respectively. In the regime $\delta < \ell \ll d$, the ADF can be
approximated by
\begin{equation}
\left\langle \Delta\Phi^2(\ell)\right\rangle_{\rm obs}
\simeq
b^2 + m^2\ell^2 + \sigma_M^2(\ell),
\end{equation}
where $b$ is the zero-separation intercept associated with the turbulent
angular dispersion, the term $m^2\ell^2$ describes the contribution
from the smoothly varying large-scale field, and $\sigma_M^2(\ell)$
accounts for the propagated measurement uncertainties of the
polarization angles.

For each stellar pair, the measurement contribution was computed from
the corresponding angular uncertainties and subsequently averaged
within each separation bin. Subtracting this contribution from the
observed ADF gives the intrinsic function
\begin{equation}
D(\ell)
=
\left\langle \Delta\Phi^2(\ell)\right\rangle_{\rm obs}
-
\sigma_M^2(\ell)
=
b^2 + m^2\ell^2.
\label{eq:adf_fit}
\end{equation}

The ADF analysis was performed independently for the NW and SE
subregions. Stars were selected according to
$P_{\rm raw}/\sigma_{P,\rm raw}\geq5$, and were
restricted to $0.3 \leq d \leq 5$~kpc. The resulting samples contain
107 stars in the NW region and 127 stars in the SE region. The
foreground Stokes parameters were determined independently for the two
subregions and subtracted from the stellar measurements prior to
computing the corrected polarization angles used in the ADF analysis.
The angular differences were computed using the $180^\circ$
periodicity of polarization orientations.

\begin{figure}
   \centering
   \includegraphics[width=.97\hsize]{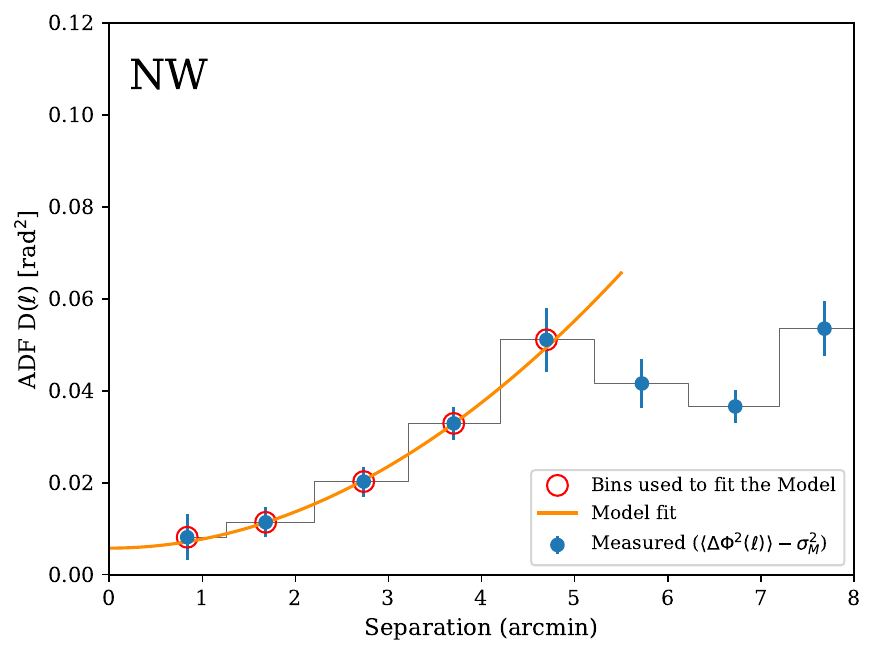}  
   \includegraphics[width=.97\hsize]{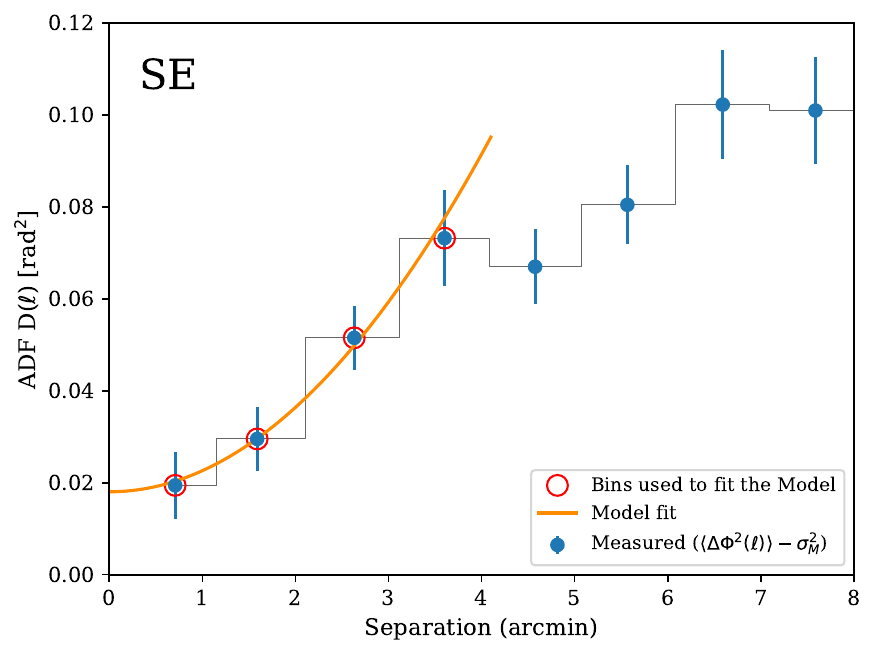} 
    \caption{Measured ADF, $( \Delta\Phi^2(\ell) - \sigma^2_\mathrm{M})$, as a 
    function of angular separation $\ell$ for the two subregions. The blue points show the measured ADF, with 
    the vertical bars indicating statistical uncertainties. The orange curves represent the best fitting ADF models, and 
    the red symbols mark the bins used in the fits. The zero-separation intercept of the fitted models yields the 
    turbulent angular dispersion parameter, $b$, used to estimate the POS magnetic-field strength. 
    At the estimated distance to the NGC\,1333 cloud, 1~arcmin corresponds to $\sim$ 0.09~pc.}
   \label{fig:adf}
\end{figure}

We constructed adaptive separation bins requiring at least 30 stellar
pair combinations and a minimum bin width of $1\arcmin$. The resulting
ADF measurements are shown in Fig.~\ref{fig:adf}. Blue symbols show
the measurement-corrected ADF, $D(\ell)$, while the orange curves
represent the best fitting linear relations in $\ell^2$. Red symbols
indicate the bins included in the final fits.

To assess the statistical stability of the fitted ADF parameters, we
performed a bootstrap at the level of the individual stars. For each
realization, the stellar catalog was resampled with replacement. If
a star appeared $c_i$ times in a given realization, an original pair
$(i,j)$ was assigned a multiplicity $c_i c_j$. The ADF was then
recomputed using these pair weights, while preserving the original
stellar geometry. We used 2000 bootstrap realizations for each
subregion.

The linear fits were performed without imposing a positivity constraint
on the intercept $b^2$. Bootstrap realizations yielding $b^2\leq0$
were therefore retained in the bootstrap analysis rather than
discarded. When expressing the bootstrap results as a physical angular
dispersion, we define
\begin{equation}
b_{\rm phys}=\sqrt{\max(b^2,0)}.
\end{equation}
The fraction of bootstrap realizations with $b^2\leq0$ was also
recorded as a diagnostic of the stability of the fitted intercept.
For the propagation of uncertainties into the DCF magnetic-field
strength, however, only the realizations with $b^2>0$ were used, since
the DCF expression contains a factor of $1/b$ and is therefore undefined
for $b=0$.

The two subregions exhibit a different dependence of the ADF on angular
scales. In the NW region, the relation remains well described by a
linear function of $\ell^2$ when extending the fit from four to five
bins. The fitted intercept changes only from $b=0.0768$~rad for four
bins to $b=0.0759$~rad for five bins, while the reduced chi-square
remains essentially unchanged, $\chi_\nu^2=0.0202$ and $0.0351$,
respectively. The five-bin fit has $\chi^2=0.105$ for three degrees
of freedom, corresponding to $p=0.991$, and the five fitted points all
remain very close to the best fitting relation. We therefore adopt the
five-bin fit for the NW region, extending the fitted ADF relation to
separations of approximately $5\arcmin$.

In contrast, the SE region is well described by the linear relation up
to separations of approximately $4\arcmin$. The four-bin fit has
$\chi^2=0.254$ for two degrees of freedom
($\chi_\nu^2=0.127$, $p=0.881$), with a fitted intercept of
$b=0.1343$~rad. Including the next separation bin increases the fitted
intercept to $b=0.1594$~rad and results in a substantially poorer fit,
with $\chi^2=6.621$ for three degrees of freedom
($\chi_\nu^2=2.207$, $p=0.085$). Although the five-bin fit is not
formally rejected at the 5\% level, the inclusion of the additional bin
modifies the fitted trend established at smaller separations. We
therefore adopt the four-bin fit for the SE region.

The zero-separation angular dispersion parameters obtained from the
full-sample fits are
\begin{equation}
b_{\rm NW}=0.0759~{\rm rad}
\end{equation}
and
\begin{equation}
b_{\rm SE}=0.1343~{\rm rad}.
\end{equation}
These full-sample values are adopted as the nominal ADF parameters.

For the bootstrap realizations with $b^2>0$, the 16th and 84th
percentiles of $b$ are $0.0443$ and $0.1115$~rad for the NW region,
and $0.0678$ and $0.1641$~rad for the SE region. Using the nominal
full-sample values as the central values, this corresponds to
\begin{equation}
b_{\rm NW}
=
0.0759^{+0.0356}_{-0.0316}~{\rm rad},
\end{equation}
and
\begin{equation}
b_{\rm SE}
=
0.1343^{+0.0298}_{-0.0665}~{\rm rad}.
\end{equation}
The fraction of bootstrap realizations with $b^2>0$ is 81.25\% in the
NW region and 96.85\% in the SE region. The corresponding fractions
with $b^2\leq0$ are 18.75\% and 3.15\%, respectively. These fractions
provide an additional indication that the uncertainty in the NW
intercept is less constrained than that in the SE region. The
bootstrap distributions are used to characterize the statistical
stability and uncertainty of the fitted parameters, while the
full-sample estimates are retained as the nominal values.

Within the ADF formalism, the turbulent-to-ordered magnetic-field
amplitude ratio is related to $b$ through
\begin{equation}
\frac{\langle B_t^2\rangle^{1/2}}{B_0}
=
\frac{b}{\sqrt{2-b^2}}.
\label{eq:Bt_B0}
\end{equation}
For the nominal full-sample values, this gives
\begin{equation}
\left(\frac{\langle B_t^2\rangle^{1/2}}{B_0}\right)_{\rm NW}
\simeq 0.054
\end{equation}
and
\begin{equation}
\left(\frac{\langle B_t^2\rangle^{1/2}}{B_0}\right)_{\rm SE}
\simeq 0.095.
\end{equation}
Applying Eq.~(\ref{eq:Bt_B0}) to the positive-$b^2$ bootstrap
realizations gives
\begin{equation}
\left(\frac{\langle B_t^2\rangle^{1/2}}{B_0}\right)_{\rm NW}
=
0.054^{+0.025}_{-0.022},
\end{equation}
and
\begin{equation}
\left(\frac{\langle B_t^2\rangle^{1/2}}{B_0}\right)_{\rm SE}
=
0.095^{+0.021}_{-0.047}.
\end{equation}

The ADF analysis reveals different angular-scale behaviors in the two
subregions. The zero-separation dispersion is larger in the SE region,
with $b_{\rm SE}=0.1343$~rad compared with $b_{\rm NW}=0.0759$~rad.
In addition, the coefficient describing the large-scale increase in
the ADF is larger in the SE, with
$m_{\rm SE}=5.42\times10^4$ compared with
$m_{\rm NW}=2.34\times10^4$. Thus, the angular dispersion increases
more rapidly with separation in the SE region. This difference is also
reflected in the range over which the linear ADF approximation is
applicable: the relation remains consistent up to $\sim5\arcmin$ in
the NW, whereas in the SE it is supported only up to $\sim4\arcmin$.
Together, these results indicate that the polarization-angle
structures sampled by the two subregions have distinct
angular-scale behavior.

\subsection{DCF Magnetic-field estimate}

The POS magnetic-field strength can be estimated through the
Davis--Chandrasekhar--Fermi (DCF) method
\citep{Davis:1951, Chandrasekhar:1953}, which relates angular
perturbations of the magnetic field to the dynamical influence of
turbulent motions. In its classical form,
\begin{equation}
B_{\rm pos}
\simeq
Q\,\sqrt{4\pi\rho}\,
\frac{\sigma_{\rm NT}}{\sigma_\theta},
\label{eq:dcf}
\end{equation}
where $\rho$ is the gas density, $\sigma_{\rm NT}$ is the
1D nonthermal velocity dispersion, and
$\sigma_\theta$ is the turbulent polarization-angle dispersion. The
factor, $Q$, typically of order $0.5$
\citep{Ostriker:2001, Crutcher:2004}, accounts approximately for
averaging effects and deviations from the idealized assumptions of the
method. We adopt $Q=0.5$.

Using
\begin{equation}
\sigma_\theta =
\frac{\langle B_t^2\rangle^{1/2}}{B_0},
\end{equation}
and Eq.~(\ref{eq:Bt_B0}), Eq.~(\ref{eq:dcf}) can be written as
\begin{equation}
B_{\rm pos}
\simeq
Q\sqrt{4\pi\rho}\,
\sigma_{\rm NT}
\frac{\sqrt{2-b^2}}{b}.
\label{eq:dcf_exact}
\end{equation}
Since the measured values of $b$ satisfy $b^2\ll2$, this expression
can be approximated by
\begin{equation}
B_{\rm pos}
\simeq
Q\sqrt{8\pi\rho}\,
\frac{\sigma_{\rm NT}}{b}.
\label{eq:dcf_m}
\end{equation}

The hydrogen column density was estimated from the visual extinction
using the gas-to-extinction conversion
\begin{equation}
N_{\rm H}=1.87\times10^{21} A_V~{\rm cm}^{-2},
\end{equation}
\citep{Bohlin:1978}. To convert the column density into a volume
density, we adopted a characteristic cloud depth of
$l=3.0\pm0.5$~pc. The corresponding mass density was calculated as
\begin{equation}
\rho=1.4m_{\rm H}n_{\rm H},
\end{equation}
where the factor of 1.4 accounts for the contribution of helium.

\begin{figure}
   \centering
   \includegraphics[width=.97\hsize]{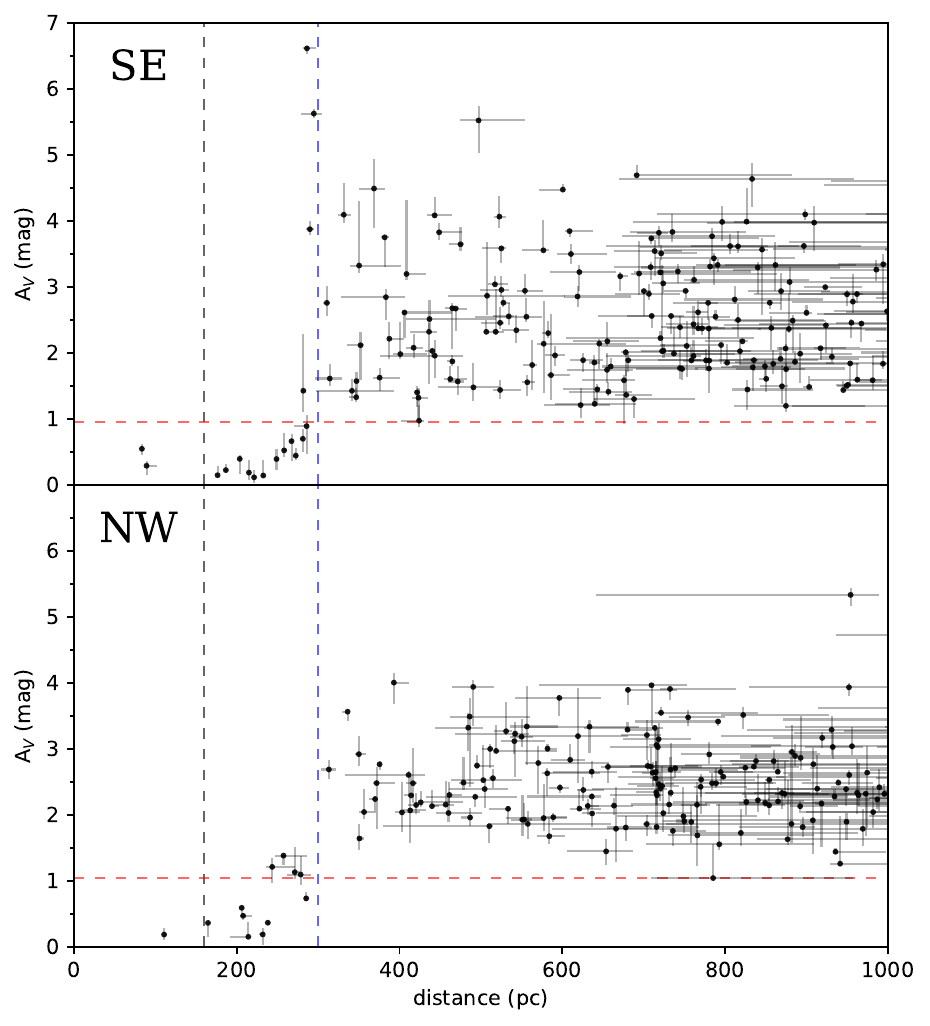}
    \caption{Visual extinction, $A_V$, as a function of distance for \texttt{StarHorse} stars projected onto the 
    SE (top) and NW (bottom) subregions. The horizontal dashed red line marks the adopted foreground extinction, 
    $A_{V, \rm NW} =1.04$ mag and $A_{V, \rm SE} =0.97$ mag, corresponding to the lower envelope of the extinction 
    distribution beyond $\sim$300 pc. The vertical dashed gray and blue lines indicate the reference distances of 160 and 
    300 pc, respectively, used to separate the nearby foreground, the intermediate layer, and the main Perseus cloud. 
    Mean cloud extinctions were derived from stars in the interval $300\leq d\leq1000$~pc after subtraction of the adopted foreground contribution.}
   \label{fig:Av_dist}
\end{figure}

To characterize the extinction distribution, we selected \texttt{StarHorse} stars
with $0.3<d_{50}\leq3.0$~kpc, $\mathrm{fidelity}\geq0.5$, and
$\mathrm{sh\_outflag}<1$, separately within the NW and SE regions.
These samples were used to inspect the extinction distribution and identify
its lower envelope. For the mean cloud extinction used in the DCF analysis,
we then considered stars in the narrower interval $300\leq d\leq1000$~pc,
as stated in the caption of Fig.~\ref{fig:Av_dist}. The distributions show
substantial sightline-to-sightline variations in extinction, reflecting the
structured distribution of dust across NGC~1333.

Rather than adopting a common foreground extinction for both
subregions, we determined the foreground contribution independently
from the lower envelope of the extinction distribution in each
region. The minimum values are
$A_{V,\rm fg}=1.04$~mag and $0.97$~mag for the NW and SE regions,
respectively. 
Subtracting these values from the corresponding mean
extinctions gives
\begin{equation}
\langle A_V\rangle_{\rm cloud,NW}
=
2.53-1.04
=
1.49~{\rm mag},
\end{equation}
and
\begin{equation}
\langle A_V\rangle_{\rm cloud,SE}
=
2.52-0.97
=
1.55~{\rm mag}.
\end{equation}

The dispersion of $A_V$ among the selected stars was interpreted as an
estimate of the sightline-to-sightline variation in the column density
across each region, rather than as the uncertainty on the mean
extinction. We therefore used the observed standard deviation of the
$A_V$ distribution in the Monte Carlo propagation. This choice
accounts for the intrinsic variation in column density sampled by the
different stellar sightlines. The resulting extinction values adopted
in the DCF analysis are therefore
\begin{equation}
A_{V,\rm NW}=1.49\pm0.63~{\rm mag}
\end{equation}
and
\begin{equation}
A_{V,\rm SE}=1.55\pm0.87~{\rm mag}.
\end{equation}

To characterize the turbulent velocity dispersion of the molecular gas
associated with NGC~1333, we considered the two $^{13}$CO velocity
components identified within the
$4 < v_{\rm LSR} < 10~\mathrm{km\,s^{-1}}$ interval (components 3 and
4; see Table~\ref{tab:co_components} and
Fig.~\ref{fig:co_lines}). For each component, the 1D
velocity dispersion was obtained from the fitted Gaussian full width at half maximum (FWHM) as
\begin{equation}
\sigma_{v,i} =
\frac{\mathrm{FWHM}_i}{\sqrt{8\ln 2}}.
\end{equation}
We then combined the dispersions of components 3 and 4 using weights
proportional to their integrated $^{13}$CO intensities,
\begin{equation}
W_i \propto
T_{{\rm peak},i}\,\mathrm{FWHM}_i,
\end{equation}
where the common Gaussian normalization factor cancels in the relative
weights. The resulting velocity dispersion was calculated as
\begin{equation}
\sigma_v^2 =
\frac{
W_3\sigma_{v,3}^2 + W_4\sigma_{v,4}^2
}{
W_3+W_4
}.
\end{equation}

Using the uncertainties of the fitted FWHMs, propagated through the
above expressions, we obtain
$\sigma_v=0.676\pm0.003~\mathrm{km\,s^{-1}}$ for the NW region and
$\sigma_v=0.908\pm0.002~\mathrm{km\,s^{-1}}$ for the SE region. We do
not include the difference between the centroid velocities of
components 3 and 4 in this quantity, since this separation represents
the relative velocity of two distinct kinematic components rather than
the internal velocity dispersion within the molecular gas. Thus, the
values adopted here characterize the internal turbulent motions of the
gas associated with NGC~1333.

The observed velocity dispersion includes both thermal and nonthermal
contributions. We therefore corrected the measured $^{13}$CO velocity
dispersion for thermal broadening according to
\begin{equation}
\sigma_{\rm NT}
=
\left(
\sigma_v^2-\sigma_{\rm th}^2
\right)^{1/2},
\end{equation}
where
\begin{equation}
\sigma_{\rm th}
=
\left(
\frac{k_{\rm B}T}{m_{^{13}{\rm CO}}}
\right)^{1/2}.
\end{equation}

Adopting a kinetic temperature of $T=15\pm5$~K
\citep{Hacar:2017}, we obtain
$\sigma_{\rm th}=0.0656^{+0.0101}_{-0.0120}~\mathrm{km\,s^{-1}}$.
Propagating the uncertainties in both the fitted velocity dispersion
and the adopted temperature gives
$\sigma_{\rm NT}\simeq0.673~\mathrm{km\,s^{-1}}$ for the NW region and
$\sigma_{\rm NT}\simeq0.906~\mathrm{km\,s^{-1}}$ for the SE region.
These nonthermal velocity dispersions are subsequently used in the
DCF analysis.

The POS magnetic-field strengths were estimated using the modified DCF
expression in Eq.~(\ref{eq:dcf_m}). Uncertainties in the ADF parameter,
extinction-derived density, cloud depth, gas velocity dispersion, and
temperature were propagated through a Monte Carlo procedure. For the
ADF parameter, the nominal values $b_{\rm fit}$ obtained from the
full-sample fits were retained as the central values, while the
bootstrap distributions with $b^2>0$ were used to characterize its
statistical uncertainty. The resulting magnetic-field parameters are
summarized in Table~\ref{tab:adf}.

The uncertainty propagation described above does not include
a separate contribution from residual sightline-to-sightline
foreground fluctuations or their spatial covariance.
Consequently, the turbulent-to-ordered magnetic-field ratios
and the POS magnetic-field strengths listed in
Table\,\ref{tab:adf} should be regarded as approximate estimates whose
uncertainties may be affected by this additional systematic
contribution.

\begin{table*}
\centering
\caption{Angular-dispersion and magnetic-field parameters for the NW
and SE regions of NGC~1333.}
\label{tab:adf}
\begin{tabular}{ccccccccccc}
\hline\hline
Region & $N_\star$ & $\ell_{\rm max}$ & $b$ &
$\langle B_t^2\rangle^{1/2}/B_0$ & $f_{b^2>0}$\tablefootmark{a}  &
$\langle A_V\rangle$\tablefootmark{b}  &
$n_{\rm H}$ & $\rho$ & $B_{\rm pos}$  \\
 & & (arcmin) & (rad) & & (\%) &(mag) &
(cm$^{-3}$) & ($10^{-22}$ g cm$^{-3}$) & ($\mu$G) \\
\hline
NW & 107 & 5 &
$0.076^{+0.036}_{-0.032}$ &
$0.054^{+0.025}_{-0.022}$ & 81.25 &
$1.49\pm0.63$ &
$300^{+147}_{-125}$ &
$7.03^{+3.47}_{-2.94}$ &
$59^{+44}_{-23}$ \\

SE & 127 & 4 &
$0.134^{+0.030}_{-0.067}$ &
$0.095^{+0.021}_{-0.047}$ & 96.85 &
$1.55\pm0.87$ &
$312^{+198}_{-154}$ &
$7.32^{+4.64}_{-3.62}$ &
$46^{+50}_{-13}$ \\
\hline
\end{tabular}
\tablefoot{
\tablefoottext{a}{$f_{b^2>0}$ denotes the fraction of the bootstrap realizations for which $b^2>0$}
\tablefoottext{b}{Corrected by the foreground component.}}
\end{table*}

\end{appendix}
\end{document}